\documentclass[12pt,preprint]{aastex}

\newcommand{\mdot}{M$_{\odot}$ yr$^{-1}$}
\newcommand{\ldot}{L$_{\odot}$}
\newcommand{ \um}{$\mu$m~}
\newcommand{ \ums}{$\mu$m}
\def\kmsMpc{\ifmmode {\rm\,km\,s^{-1}\,Mpc^{-1}}\else
    ${\rm\,km\,s^{-1}\,Mpc^{-1}}$\fi}

\shorttitle{Infrared Spectra and SEDs of Dusty Galaxies}
\shortauthors{Sargsyan et al.}

\begin{document}

\title{Infrared Spectra and Spectral Energy Distributions for Dusty Starbursts and AGN}

\author{Lusine Sargsyan\altaffilmark{1}, Daniel Weedman\altaffilmark{2}, Vianney Lebouteiller\altaffilmark{2,3}, James Houck\altaffilmark{2}, Donald Barry\altaffilmark{2}, Ashot Hovhannisyan\altaffilmark{4} and Areg Mickaelian\altaffilmark{1} }
\altaffiltext{1}{Byurakan Astrophysical Observatory, 378433, Byurakan, Aragatzotn Province, Armenia; sarl11@yahoo.com}
\altaffiltext {2}{Astronomy Department, Cornell University, Ithaca,
NY 14853; dweedman@isc.astro.cornell.edu}
\altaffiltext {3} {Laboratoire AIM, CEA/DSM-CNRS-Universite Paris Diderot, DAPNIA/Service d'Astrophysique, Saclay, France}
\altaffiltext {4} {Department of Physics, Yerevan State University, Yerevan, Armenia}

\begin{abstract}
  
We present spectroscopic results for all galaxies observed with the $Spitzer$ Infrared Spectrograph (IRS) which also have total infrared fluxes $f_{IR}$ measured with the Infrared Astronomical Satellite (IRAS), also using AKARI photometry when available.  Infrared luminosities and spectral energy distributions (SEDs) from 8 \um to 160 \um are compared to polycyclic aromatic hydrocarbon (PAH) emission from starburst galaxies or mid-infrared dust continuum from AGN at rest frame wavelengths $\sim$ 8 \ums.  A total of 301 spectra are analyzed for which IRS and IRAS include the same unresolved source, as measured by the ratio $f_{\nu}$(IRAS 25 \ums)/$f_{\nu}$(IRS 25 \ums).  Sources have 0.004 $<$ z $<$ 0.34 and 42.5 $<$ log $L_{IR}$ $<$ 46.8 (erg s$^{-1}$) and cover the full range of starburst galaxy and AGN classifications.  Individual spectra are provided electronically, but averages and dispersions are presented.  We find that log [$L_{IR}$/$\nu L_{\nu}$(7.7 \ums)] = 0.74 $\pm$ 0.18 in starbursts, that log [$L_{IR}$/$\nu L_{\nu}$(7.7 \ums)] = 0.96 $\pm$ 0.26 in composite sources (starburst plus AGN), that log [$L_{IR}$/$\nu L_{\nu}$(7.9 \ums)] = 0.80 $\pm$ 0.25 in AGN with silicate absorption, and log [$L_{IR}$/$\nu L_{\nu}$(7.9 \ums)] = 0.51 $\pm$ 0.21 in AGN with silicate emission.  $L_{IR}$ for the most luminous absorption and emission AGN are similar and 2.5 times larger than for the most luminous starbursts.  AGN have systematically flatter SEDs than starbursts or composites, but their dispersion in SEDs overlaps starbursts.  Sources with the strongest far-infrared luminosity from cool dust components are composite sources, indicating that these sources may contain the most obscured starbursts.

\end{abstract}


\section{Introduction}

Infrared observations are crucial to understanding the formation and evolution of galaxies, because the most luminous galaxies in the universe are dusty and heavily obscured \citep*{soi87}.  Most of their primary luminosity arising from optical and ultraviolet luminosity is absorbed and subsequently reradiated in the infrared by the absorbing dust.  Three decades of discovery with the Infrared Astronomical Satellite (IRAS), the Infrared Space Observatory (ISO), and the $Spitzer$ Infrared Observatory ($Spitzer$) have traced evolution for the most luminous galaxies in the universe to redshift z $\sim$ 3 (brief review in \citet{hou10}).  

The epoch with 2 $\la$ z $\la$ 3 is distinctive because it is the epoch at which the luminosity of optically-discovered sources is observed to peak \citep [e.g.][]{mad98, red09}.  It is important to characterize thoroughly all populations found within this epoch.   Once that is accomplished, the next frontier is to understand the initial formation of galaxies by studying sources at redshifts z $>$ 3. When were the first galaxies with rapid and intense star formation ("starbursts") and the first active galactic nuclei (AGN)?  What fraction of these luminous sources are obscured by dust?  When did dust emission become the dominant luminosity in the universe? 

For example, it is essential to determine whether optically-discovered sources to z $\sim$ 8 observed in the rest-frame ultraviolet \citep [e.g.][]{bou09} are the dominant population in the early universe.  Dusty galaxies would not be found within such optically selected samples at high redshift because of the very large extinction in the rest frame ultraviolet.  Comparison of starburst luminosity in the mid-infrared and ultraviolet rest frames indicates that less than 1\% of the intrinsic ultraviolet luminosity emerges from the most luminous infrared-selected sources \citep [e.g.][]{sar10}.  Does this very large dust obscuration apply to sources in the early universe and prevent optical discovery of the most luminous galaxies?  Answers to such questions will be determined by future observations at far-infrared to millimeter wavelengths, where dusty galaxies can be discovered at redshifts z $>$ 3. 

Mid-infrared spectra are available from the $Spitzer$ Infrared Spectrograph (IRS; \citet{hou04}) for hundreds of optically-faint infrared sources, including starbursts, AGN, and composite sources with signatures of both.  These sources range in luminosity from 10$^{8}$ to 10$^{14}$ \ldot~ and in redshift from 0 $<$ z $<$ 3.  These sources were discovered in surveys at 24 \um with the $Spitzer$ Multiband Imaging Photometer System (MIPS; \citet{rie04}) and subsequently observed with the IRS. 

Results and source lists for dusty starbursts and AGN are summarized in \citet {wee08} and \citet {wee09b}.  (Because the objective of these summaries was to determine separately the characteristics of starbursts and AGN, composite sources were not included.)  Primary sources for spectroscopic results are \citet {hou05, hao05, bra06, shi06, wee06a, wee06b, arm07, far07, hou07, ima07, mk07, saj07, sch07, yan07, brn08a, brn08b, far08, mar08, pol08, pop08, sar08, das09, des09, far09a, her09, hua09, men09, wee09a} and \citet{wee09b}.  

These sources provide the available sample of dusty, luminous galaxies extending to z $\sim$ 3.  To interpret these sources and relate them to future new discoveries, it is necessary to determine how the mid-infrared spectroscopic parameters defining classification and luminosities of starbursts and AGN relate to total infrared luminosities and photometric parameters derived from spectral energy distributions (SEDs).  The purpose of the present paper is to compare mid-infrared spectroscopic features with total infrared luminosities and spectral energy distributions (SEDs) for large numbers of sources having available both mid-infrared $Spitzer$ IRS spectra and also far infrared photometric results.  

This comparison will allow:  determining luminosities from mid-infrared spectroscopic parameters; relating local starbursts and AGN to far infrared and submillimeter  sources discovered at high redshifts; determining parameters which can distinguish starbursts and AGN, particularly the long wavelength, cold dust component;  and determining empirical templates and their dispersion among starbursts and AGN for modeling source counts and source colors within surveys at various far infrared to millimeter wavelengths.

We have two primary objectives for this paper.  The first is to determine the total luminosities of sources as a function of mid-infrared spectral features, so that the large numbers of $Spitzer$ spectra available for sources with z $<$ 3 can be used to determine the total luminosities of starbursts and AGN at these epochs.  Our second primary objective is to utilize a consistent mid-infrared spectral classification for sorting objects among starbursts, AGN, and composites, and then to determine median SEDs of these different categories.  This will allow a conclusion of how accurately SEDs alone can classify high redshift, luminous sources which are being discovered with far infrared and submillimeter observations. 

The spectral features we use for our analysis are simply the localized spectral maxima arising from peak flux densities for the polycyclic aromatic hydrocarbon (PAH) 7.7 \um emission feature in spectra of sources containing starbursts, and the 7.9 \um spectral continuum peak for obscured AGN with silicate absorption. (For unobscured AGN with neither feature, the spectral continuum at 7.9 \um is measured.)  Examples are illustrated in Figure 1 to show how such features would appear in observed IRS spectra of sources such as those analysed below if redshifted to z = 2.5.  Our objective is to measure the strongest features, motivated by using these parameters to have reliable and self-consistent data for the large numbers of sources discovered by $Spitzer$ at z $\ga$ 2 for which S/N is poor, and deconvolution of the total flux within spectral features is uncertain \citep{hou07,wee08,wee09a,wee09b}.  

Use of these spectral features will allow comparisons of SEDs and source luminosities for our local sample to the samples at z $\ga$ 2 with available spectra.  These new comparisons can be made because the high redshift sources having $Spitzer$ spectra were originally discovered within $Spitzer$ survey fields that are now included in mapping to wavelengths of 500 \um in the Herschel Multi-tired Extragalactic Survey (HerMES; \citet{oli10}). SEDs for these high redshift sources can be compared at the same far-infrared rest frame wavelengths included in our local sample \citep [e.g.][]{rr10,cha10}.
 
To achieve our objectives, we assemble and analyze a collection of 301 sources which have both IRS spectra and IRAS photometry;  when available, photometry extending to 160 \um is also used from the AKARI mission \citep{mur07,ona07,kaw07,kan07}.  Using this sample, the spectral features from $Spitzer$ IRS spectra at rest frame $\sim$ 8 \um are compared to photometric parameters from 8 \um to 160 \um measured with IRAS and AKARI.  

Individual spectra of the sources which we discuss in this paper are available electronically. \footnote{Query on source name at http://cassis.astro.cornell.edu/sargsyan-etal; The Cornell Atlas of Spitzer IRS sources (CASSIS; V. Lebouteiller et al. in preparation) is a product of the Infrared Science Center at Cornell University, supported by NASA and JPL.}



\section{Sample Selection and Observations}

All sources were observed with the IRS using the low resolution modules Short Low 1 (SL1), Short Low 2 (SL2), Long Low 1 (LL1) and Long Low 2 (LL2), giving spectral coverage from $\sim$5\,\um to $\sim$35\,\ums.  For all spectra which we analyzed, spectral extractions were done with the SMART analysis package \citep{hig04}, using the improved "optimal extraction" procedure in \citet{leb10}\footnote{http://isc.astro.cornell.edu/IRS/SmartDoc}.  

The optimal extraction procedure fits an empirical point spread function (PSF) for an unresolved source to the spectral image and weights each pixel in the image according to its fraction of illumination by the source.  This reduces the noise arising from the background.  Spectral flux calibration is provided using observations of standard stars; this calibration corrects for slit losses for unresolved sources.  The flux calibration is correct to within $\sim$ 5\% for unresolved sources.  For resolved sources, the fitting of the PSF and finite slit size results in an underestimate of the flux density in the extracted spectrum.   

Our objective is to compare spectral features from the IRS with overall spectral energy distributions (SEDs) determined photometrically to far infrared wavelengths using IRAS and AKARI.  To make such comparisons, it is necessary to use only unresolved sources, so that different spatial resolutions do not affect the comparisons.  Our estimate of source extension arises by comparing flux densities measured through the IRS slit with total flux densities of sources measured photometrically by IRAS.  This is the criterion we have used to restrict our sample to unresolved sources, as now described.  

We first examined the IRS archive of $Spitzer$, selecting sources having both IRS low resolution spectra and complete flux measures with IRAS. This yielded a sample of 501 sources.  For these sources, we extracted all IRS LL1 spectra and determined the ratio of synthetic IRAS 25 \um flux density measured on the IRS spectrum, $f_{\nu}$(IRS 25 \ums), compared to the $f_{\nu}$(IRAS 25 \ums) observed with IRAS.  Comparing the total $f_{\nu}$(IRAS 25 \ums) with the $f_{\nu}$(IRS 25 \ums) measured with the 10.7\arcsec~ slit of the IRS LL1 gives a measure of source extension.  The resulting distribution is shown in Figure 2.  

Figure 2 shows a concentration of sources near a value of $f_{\nu}$(IRAS 25 \ums)/$f_{\nu}$(IRS 25 \ums) of unity, as expected for unresolved sources.   There is a dispersion of $\sim$ 20\% for the core distribution near unity.   This dispersion reflects uncertainties in the comparative flux calibrations of the IRS and IRAS and in the synthetic photometry, because real values below unity cannnot occur (unless sources are actually variable).   

To accomodate the dispersion arising from such uncertainties while minimizing uncertainties arising from aperture corrections for extended sources, we selected from the initial sample of 501 objects only those objects having flux ratios $f_{\nu}$(IRAS 25 \ums)/$f_{\nu}$(IRS 25 \ums) $<$ 1.5.  This results in a final sample of 301 objects.  This selection criterion assures that the final sample of 301 sources can be considered as spatially unresolved sources at infrared wavelengths longer than $\sim$ 20 \um, so that the IRS spectral parameters refer to the same source as the longer wavelength SEDs from IRAS.  

We do not apply corrections for slight source extension by using values of $f_{\nu}$(IRAS 25 \ums)/$f_{\nu}$(IRS 25 \ums).  This is partly because much of the differences from unity can be attributed to flux uncertainties.  It is also because slightly resolved sources would not necessarily have the same spectra and SEDs throughout the source, so the true corrections are unknown.  The resulting overall uncertainties in comparing IRS spectral parameters to IRAS and AKARI photometry are reflected in the empirical dispersions for these relations discussed in sections 3.2 and 3.4.  All spectroscopic and photometric data for our final sample are listed in Tables 1-6. 

The primary spectral parameter used in our analysis is at $\sim$ 8 \um, which is observed with the IRS SL module, having a narrower slit (3.7\arcsec) compared to LL.  In a few cases, sources are resolved at this scale, which is noted in a lack of overlap between the SL and LL portions of the spectrum.  In such cases, the SL flux densities are increased by the factor needed to overlap LL (the "stitching" factor).  

To measure total infrared luminosity for sources, we use $L_{IR}$ defined as in \citet{san96}, with $f_{IR}$ =  1.8 x 10$^{-11}$[13.48$f_{\nu}$(12) + 5.16$f_{\nu}$(25) + 2.58$f_{\nu}$(60) + $f_{\nu}$(100)], for $f_{IR}$ in erg cm$^{-2}$ s$^{-1}$ and IRAS flux densities $f_{\nu}$(12), $f_{\nu}$(25), $f_{\nu}$(60), and $f_{\nu}$(100) in Jy. This relation includes an estimate for additional long wavelength flux beyond the 100 \um limit of IRAS. 

$L_{IR}$ is a measure of total luminosity which is reradiated by absorbing dust.  When sources are heavily obscured in all directions, $L_{IR}$ is also a measure of bolometric luminosity, because primary radiation from all shorter wavelengths is always absorbed and reemitted from the dust.  If absorption by dust is not uniform over all directions, as in AGN having increased absorption within a dusty torus, $L_{IR}$ underestimates the true bolometric luminosity. 

Our sample is representative of fainter, flux limited infrared samples chosen with IRAS or $Spitzer$.  This is illustrated in Figure 3 by comparing mid infrared $f_{\nu}$(25 \ums) with near infrared $f_{\nu}$(1.2 \ums) from the Two Micron All Sky Survey (2MASS; \citet{skr06}).  This comparison shows the ratio of dust continuum to continuum from stars or non-thermal AGN continuum.  Results indicate that the present sample spans the full range of dust obscuration and dust emission found in any sample of infrared-discovered sources brighter than 10 mJy at 25 \ums.

\subsection{Infrared and Optical Spectral Classifications}

We desire to relate the local objects in the present sample to spectroscopic classifications for dusty sources at z $\sim$ 2.  Our primary spectroscopic classification, therefore, is a classification of starburst, AGN, or composite (starburst plus AGN) derived from spectra at rest frame wavelengths between approximately 5 \um and 12 \um because these are the rest frame wavelengths observed with the IRS for z $\ga$ 2 (Figure 1).  

The single most important parameter in these IRS spectra for a mid-infrared measure of the starburst component is the flux in PAH molecular emission features \citep{gen98,bra06,hou07,wee09a,sar09,far09b}. The photodissociation region at the interface between the surrounding molecular cloud and the HII region of a starburst provides the appropriate physical conditions for strong PAH excitation and emission \citep{pee04}.  Our spectral classification for starbursts, AGN, and composites depends only on the strength of these PAH features.

Figure 4 shows the total infrared luminosity $L_{IR}$ determined using IRAS fluxes compared with the rest-frame equivalent width of the 6.2 \um PAH feature for our sample of sources.  The EW(6.2 \ums) determines the fraction of starburst (SB) compared to AGN luminosity; the divisions into AGN, Composite and SB which we adopt are illustrated in Figure 4.  

These divisions derive from previous studies with the IRS of sources having optical classifications.  The definition of "pure starburst", with EW(6.2 \ums) $>$ 0.4 um, arose originally from empirical measures of infrared spectra from the sample of optically classified starbursts in \citet {bra06}; this EW value was subsequently found to divide starbursts and AGN in infrared flux-limited samples \citep{wee09a,wu10}.  Figure 4 shows the comprehensive range of our sample in both AGN/starburst fraction and in luminosity.  

Perfect agreement between infrared and optical classifications is not expected for these dusty, obscured sources in which the primary luminosity source is often completely hidden optically. There are also well known examples of optical AGN with circumnuclear starbursts, so either might dominate the infrared spectrum.  Figure 4 nevertheless shows the general consistency between infrared and optical classifications, using classifications derived from optical spectra with the Sloan Digital Sky Survey (SDSS; \citet{gun98}) given in Tables 1, 3, and 5.  

Sources are separated in these Tables according to the classifications of EW(6.2 \micron) in Figure 4.  For starbursts in Table 1, optical spectra when available show indications of an AGN in only one of 19 sources classified as starburst using the infrared criterion; the one is a composite classification. Composite sources in Table 3 show a mix of optical classifications, as expected.  Using the limited number of AGN in Table 5 which have optical spectral classifications, optical spectra when available show the presence of an AGN (AGN or composite optical classification) in 50 of 56 sources classified as AGN using the infrared criterion. 

The spectral features at rest frame wavelengths $\sim$ 8 \um (Figure 1) which we use to make uniform measures of source luminosity can be traced with the IRS through all redshifts for 0 $<$ z $\la$ 3.  For starbursts, the spectral peak is at 7.7 \um and arises from PAH emission; for absorbed AGN, this peak is not an emission feature but is a continuum maximum between absorption features and is typically at 7.9 \ums.  In all of our measurements and discussion for this paper, we measure the spectral peak whenever such a peak is present, and we refer in general to such measurements of the peak as $f_{\nu}$($\sim$ 8 \ums).  For silicate emission AGN without either detectable PAH at 7.7 \um or a localized continuum maximum, our measures are of the continuum at 7.9 \ums, which we also define as $f_{\nu}$($\sim$ 8 \ums).

Our measure of EW(6.2 \ums) makes the assumption that this feature is a Gaussian on an underlying continuum which is the continuum immediately adjacent to the Gaussian, the standard definition of EW.  We use this measure of EW to provide a consistent parameter among various sources, even those with poor S/N.  In reality, the PAH features are broad and complex, and the 6.2 \um feature rests on wings of other PAH features, so that the adjacent continuum is not the true underlying continuum from dust emission. The complex of PAH features can be deconvolved if assumptions are made regarding relative fluxes and the spectral distribution of the underlying dust continuum \citep{smi07}.  An accurate deconvolution is not feasible in the spectra of faint sources at high redshift which have poor S/N and limited spectral coverage; this is why we have introduced the peak $\nu$$f_{\nu}$(7.7 \ums) as a measure of PAH luminosity. 

To demonstrate that the peak $\nu$$f_{\nu}$(7.7 \ums) is equivalent to measuring the total flux within an individual PAH feature for starbursts (and only starbursts), we compare in Figure 5 the $\nu$$f_{\nu}$($\sim$ 8 \ums) for all sources with the total flux measured in the PAH 6.2 \um feature.  For starbursts, most composites, and many AGN, $f_{\nu}$($\sim$ 8 \ums) is the peak of the 7.7 \um PAH feature; for AGN and a few composites without a 7.7 \um PAH feature, $f_{\nu}$($\sim$ 8 \ums) is the continuum at 7.9 \ums. Figure 5 shows, as expected, that weaker PAH measured by EW corresponds to an increased underlying dust continuum which artificially enhances the peak $\nu$$f_{\nu}$($\sim$ 8 \ums).  An increasing contribution to the dust continuum at 8 \um from an AGN which underlies the PAH feature from a starburst would explain all of the trends seen in this plot.  

For the sources we classify as "pure" starbursts, with EW(6.2 \ums) $>$ 0.4 \ums, the median and 1 $\sigma$ dispersion in Figure 5 for the ratio $\nu$$f_{\nu}$(7.7 \ums)/1000 $f$(6.2 \ums) is 0.065 $\pm$ 0.006.  This empirical result indicates a 10 \% uncertainty in use of the peak $\nu$$f_{\nu}$(7.7 \ums) to measure PAH luminosity from a starburst when compared to a measure of total luminosity in individual PAH features.  This uncertainty is a measure of varying levels for the true dust continuum underlying the PAH features, but this is not an uncertainty in the luminosity of the starburst as measured by $\nu$$f_{\nu}$(7.7 \ums) if the underlying continuum also arises from dust heated only by the starburst.

\section{Discussion}

\subsection{Total Infrared Fluxes Compared to Spectral Parameters}

Our previous summaries determined luminosities of dusty, obscured sources discovered with $Spitzer$ using $\nu L_{\nu}$($\sim$ 8 \ums), and showed that the most luminous starbursts and AGN show luminosity evolution in this parameter of form (1+z)$^{2.5}$ to z $\sim$ 2.5 \citep{wee08, wee09b}.  Transforming $\nu L_{\nu}$($\sim$ 8 \ums) to $L_{IR}$ is a crucial calibration for determining star formation rates (SFR) derived from total infrared luminosities as in \citet{ken98}, and for measuring total luminosities of AGN.  Our initial calibrations for sources to z $\sim$ 2.5 were based on small samples and need improvement to determine how the calibrations depend on starburst/AGN fraction and on luminosity.  Improving this calibration is the objective of this section. 

Using sources classified in Tables 1, 3, and 5 as "pure" starbursts [EW(6.2 \um) $>$ 0.4 \um)], composite sources [0.1 \um $<$ EW(6.2 \um) $<$ 0.4 \um], and "pure" AGN [EW(6.2 \um) $<$ 0.1 \um], we show in Figure 6 the ratios $f_{IR}$/$\nu$$f_{\nu}$($\sim$ 8 \ums) for all sources.  It is these ratios that allow estimates of total infrared luminosities $L_{IR}$ when only mid-infrared spectra are available. 

For starbursts in Figure 6, the median log [$f_{IR}$/$\nu$$f_{\nu}$(7.7 \ums)] = 0.74 $\pm$ 0.18.  (A value of 0.78 was used in \citet{wee08}.)

For most composite sources, the PAH features are sufficiently strong that the spectral peak is at 7.7 \ums, but for some sources with weak PAH and silicate absorption, the peak is at 7.9 \ums, as noted in Table 3.  In Figure 6, the median log [$f_{IR}$/$\nu$$f_{\nu}$(7.7 \um or 7.9 \ums)] = 0.96 $\pm$ 0.26.  (Composite sources were not used in our previous analyses of "pure" starbursts or "pure" AGN.) 

For AGN, there is sometimes a sufficiently strong 7.7 \um PAH feature that this is the spectral peak.  The spectral peak is at $\sim$ 7.9 \um if the source has silicate absorption.  If the source has no localized peak near 8 \ums, either from PAH emission or silicate absorption, our spectral measurement is made at a wavelength of 7.9 \ums.  Which parameter is measured is noted in Table 5.  All AGN are combined in Figure 6 for comparison to starbursts and composite sources, but the $f_{IR}$/$\nu$$f_{\nu}$(7.7 \um or 7.9 \ums) is determined separately for silicate absorption and silicate emission AGN as shown in Figure 7. 

For silicate emission AGN in Figure 7, the median log [$f_{IR}$/$\nu$$f_{\nu}$($\sim$ 8 \ums)] = 0.51 $\pm$ 0.21. (A value of 0.74 was used in \citet{wee09b}.)

For silicate absorption AGN in Figure 7, the median log [$f_{IR}$/$\nu$$f_{\nu}$($\sim$ 8 \ums)] = 0.80 $\pm$ 0.25. (A value of 0.95 was used in \citet{wee09b}.)
Although this is the median and dispersion for all absorbed AGN,  Figure 7 indicates that the most strongly absorbed sources have systematically larger values of $f_{IR}$/$\nu$$f_{\nu}$($\sim$ 8 \ums).  This can be explained if the absorption also affects the continuum at $\sim$ 8 um, such that luminosity removed from the mid-infrared reappears in the far infrared. This is consistent with an extinction correction discussed more below. 

That the largest values of $f_{IR}$/$\nu$$f_{\nu}$($\sim$ 8 \ums) arise for composite sources is an interesting result from Figure 6.  The composite sources are defined by intermediate strength of the PAH features.  If composites are any combination of AGN and starbursts such as those included in the AGN and starburst samples, the composites should have $f_{IR}$/$\nu$$f_{\nu}$($\sim$ 8 \ums) within the range of these ratios for AGN and starbursts.  But the composite ratio exceeds both.

There is an explanation for the composite sources which can explain this result.  This would arise if the "composite" classification includes sources with PAH which is weak because the PAH from the starbursts in these sources are more obscured compared to the PAH within the "starburst" classification.  In this circumstance, the PAH would be weakened by absorption, and the absorbed luminosity would reappear at longer wavelengths.  This would produce larger values of $f_{IR}$/$\nu$$f_{\nu}$($\sim$ 8 \ums).  Adopting this explanation implies that that sources with the greatest far-infrared luminosity and the coolest dust are sources at the extreme of obscuration for starbursts.  This question is considered again in section 3.4 in context of SEDs.

\subsection{ Luminosities of the Most Luminous Starbursts and AGN}

Using all available IRS observations, \citet{wee08, wee09b}, and \citet{hou10} fit luminosity evolution for 0 $<$ z $<$ 2.5 of the most luminous starbursts and absorbed AGN as measured by $\nu L_{\nu}$($\sim$ 8 \ums).  This is the parameter used in the present paper for comparison to  $L_{IR}$, so that luminosity evolution in $L_{IR}$ can be determined using the ratios $f_{IR}$/$\nu$$f_{\nu}$($\sim$ 8 \ums) determined above.

Using our new calibrations together with these forms of evolution, we can determine the total luminosity $L_{IR}$ as a function of redshift for the most luminous dusty starbursts and AGN. Results are: 

\noindent a) For the most luminous starbursts,
\begin{equation}
$$log $L_{IR}$(SB) = 11.8$\pm$0.18 + 2.5($\pm$0.3) log(1+z) for $L_{IR}$ in \ldot.$$ 
\end{equation} 

An important use of $L_{IR}$(SB) is to determine star formation rate (SFR) from the precepts of \citet{ken98}, for which log (SFR) = log $L_{IR}$ - 9.76, for SFR in units of \mdot and $L_{IR}$ in \ldot.   


\noindent b) For the most luminous obscured, silicate absorption AGN, 
\begin{equation}
$$log $L_{IR}$(AGN$_{obscured}$) = 12.2$\pm$0.25 + 2.6($\pm$0.3) log(1+z) for $L_{IR}$ in \ldot.$$ 
\end{equation} 

In the "unified model", the only difference between emission and absorption AGN is the viewing angle.  Our results are consistent with this model such that all AGN have intrinsically similar total infrared luminosity, but absorbed AGN appear fainter in mid-infrared wavelengths because of absorption at those wavelengths.  The absorbed luminosity from the mid-infrared spectrum reappears at longer wavelengths.  We have found in section 3.1 that the median $f_{IR}$/$\nu$$f_{\nu}$($\sim$ 8 \ums) = 6.5 for absorption sources and 3.2 for emission sources, as shown in Figure 7.  This implies a factor of two extinction correction is required for $f_{\nu}$($\sim$ 8 \ums) for the absorbed sources. 

Because of insufficient sources in the various redshift bins, \citet{wee09b} did not fit the form of evolution for silicate emission sources.  The luminosity distribution $\nu L_{\nu}$($\sim$ 8 \ums) with redshift for silicate emission sources is updated in \citet{hou10}.  This distribution has similar shape to the distribution of absorbed sources, except that the emission sources are offset to brighter $\nu L_{\nu}$($\sim$ 8 \ums) luminosities.  This offset is consistent with the factor of two which would arise from the extinction correction for the absorbed sources.  We conclude, therefore, that $L_{IR}$ for the most luminous emission sources is the same as for the most luminous absorbed sources.  With this conclusion, the luminosity evolution for the most luminous unobscured, silicate emission, type 1 AGN also becomes: 
\begin{equation}
$$log $L_{IR}$(AGN1) = 12.2$\pm$0.21 + 2.6($\pm$0.3)log(1+z) for $L_{IR}$ in \ldot.$$ 
\end{equation} 

These results are similar to those in \citet{wee09b} and \citet{hou10}, but the log $L_{IR}$ are systematically less luminous because of changes in the 
$f_{IR}$/$\nu$$f_{\nu}$($\sim$ 8 \ums) calibration.  For starbursts, log $L_{IR}$(SB) decreases by 0.04; for silicate absorption AGN, log $L_{IR}$(AGN$_{obscured}$) decreases by 0.14; and for silicate emission AGN, log $L_{IR}$(AGN1) decreases by 0.23 compared to our earlier conclusions.   

These three equations indicate that, within uncertainties, the total infrared luminosities of the most luminous starbursts and most luminous AGN are similar to within a factor $\sim$ 2.  This result is also consistent with observed spectra representing the brightest examples shown in Figure 1.  The brightest sources observed at z = 2.5 in rest-frame $f_{\nu}$($\sim$ 8 \ums) scale as illustrated, with $f_{\nu}$(7.9 $\mu$m, AGN1) : $f_{\nu}$(7.9 $\mu$m, AGN$_{obscured}$) : $f_{\nu}$(7.7 $\mu$m, SB) = 12 : 5 : 2, in mJy.  Applying our $f_{IR}$/$\nu$$f_{\nu}$($\sim$ 8 \ums) calibrations to these values of $f_{\nu}$($\sim$ 8 \ums) then yields $L_{IR}$(AGN1) : $L_{IR}$(AGN$_{obscured}$) : $L_{IR}$(SB) = 3.5 : 2.9 : 1.0 for these individual brightest examples.  Scaling from the formal equations 1-3 would give the similar result that $L_{IR}$(AGN1) : $L_{IR}$(AGN$_{obscured}$) : $L_{IR}$(SB) = 2.5 : 2.5 : 1.0. 

It is important to note, however, that $L_{IR}$(AGN)/$L_{IR}$(SB) probably underestimates the value of this ratio for total bolometric luminosities.  If AGN luminosity is absorbed only within an obscuring torus where absorbing dust produces $L_{IR}$(AGN), some primary radiation would escape the torus without being absorbed.  This escaping radiation is not included in our measure of $L_{IR}$(AGN).  By contrast, starbursts arising within the obscuring torus would have all of their bolometric luminosity absorbed and reradiated by the dust. 

Our previous analyses which determined these forms of luminosity evolution were separated for sources classified as predominantly starbursts or predominantly AGN.  This was done in an effort to determine if evolution appears different for these categories.   The similarity of the results shows that luminosity evolution occurs together for starbursts and AGN as classes of sources.  Whatever process triggers the formation of luminous, dusty galaxies triggers both starbursts and AGN.  This conclusion does not, however, determine for an individual galaxy whether there is evolution between starburst and AGN.  Descriptions of such "individual evolution" from one class to another \citep [e.g.][]{san96,spo07,far09b} are consistent with the similarity of overall luminosity evolution which we find for both classes.   

In our view, the most important question regarding individual evolution is whether AGN or starbursts came first in the early universe.   We now have extensive samples of both to redshifts approaching 3, with no evidence of differences in evolution for these different classes of luminous infrared sources.   Answering the crucial questions will require discovery and classification of luminous, dusty galaxies at higher redshifts.

\subsection {Shapes of Spectral Energy Distributions}

The $Spitzer$ IRS results summarized above trace evolution of luminous, dusty galaxies to z $\sim$ 2.5.  To follow such sources beyond such redshifts requires new far-infrared, submillimeter, and millimeter observations.  Redshifts may not be available for such sources, other than estimates of photometric redshifts based on SED shapes.  In this section, we determine empirically the median SEDs and dispersions of the sources classified spectroscopically.  Our objective is to understand if SEDs are sufficiently different that sources can be classified with long wavelength photometric observations using only the SEDs. 

To extend SEDs to longer wavelengths than available from IRAS, we use AKARI fluxes to 160 \um when available\footnote{IRAS flux densities were obtained from the IRAS Faint Source Catalog at http://vizier.u-strasbg.fr/viz-bin/VizieR-4. AKARI flux densities are from the AKARI catalog at http://darts.isas.jaxa.jp/astro/akari/cas/tools/search/crossid.html.}.  The most important use of these AKARI results is to consider the coolest dust components, as discussed below and illustrated in Figure 13.  We note that there seem to be systematic differences between IRAS and AKARI photometric results at the level of $\sim$ 10\%.  These differences are shown in Figure 8.  

In Figure 8, we show the median results for all 190 objects from our total sample of 301 that have complete AKARI flux density measures.  This figure shows that the median SED is not smooth between 60 \um and 100 \um where IRAS and AKARI measures overlap.  Discontinuities are present at the $\sim$ 10\% level for flux densities between 60 \um and 100 \um.  These discontinuities will also appear in our median SEDs for different classifications (starbursts, composites, and AGN) discussed below.  We do not believe that these are real discontinuities in the SEDs and conclude that these arise from photometric uncertainties.  These uncertainties are small compared to the observed dispersions among SEDs, so we make no effort at reconciling these differences.  Refinements can be made when new photometry is available from the Herschel Observatory \citep{pil10,gri10}.

The distribution of SEDs for all starburst sources as determined from IRAS and AKARI flux densities, in the rest frame normalized to the peak $\nu$$f_{\nu}$(7.7 \ums), is shown in Figure 9. Solid lines show medians at the various observing wavelengths, and error bars show one $\sigma$ dispersions at medians of the various rest frame wavelengths observed by IRAS and AKARI. The distribution of SEDs for all composite sources as determined from IRAS and AKARI flux densities, in the rest frame normalized to the peak $\nu$$f_{\nu}$(7.7 \um or 7.9 \ums), is shown in Figure 10, and the distribution of SEDs for all AGN is shown in Figure 11. The dispersions among the SEDs are compared in Figure 12.  There are several notable results from these comparisons.

The "pure" starbursts show the most consistent spectra at all wavelengths.  The dispersions generally range only over a factor of two at any wavelength.   This means that a mid-infrared classification as starburst is a good predictor of far infrared properties.  

As known since IRAS, typical SEDs for AGN are flatter than for starbursts, because AGNs contain a component of hotter dust that raises the mid-infrared dust continuum compared to the far infrared.  This result is evident in Figure 12.  When considering the dispersion among sources, however, the AGN are not invariably distinct from starbursts.   The dispersions shown in Figure 12 overlap.  There are many objects classified purely as AGN in the mid-infrared spectrum which have cool dust components as significant as in starbursts.   As already commented above in section 3.1, the coolest dust components (greatest far infrared luminosities) are found in the composite sources, those with weak mid-infrared indicators of a starburst.

What is the source of the cool dust component in AGN and in the composite sources?  In context of dusty torus models for AGN, it has been suggested that the cool dust is also heated by the AGN \citep [e.g.][]{pol08} but is cool either because it is far from the AGN or is behind optically thick dust.  A more common assumption is that cool dust is invariably a starburst indicator \citep [e.g.][]{far09b}; in this case, the absence of starburst signatures in the mid-infrared for composite sources and AGN would be attributed to very heavy extinction of the starburst luminosity.  

Despite this ambiguity in the far-infrared SEDs,  Figures 9-11 illustrate that there are limits to the shape of SEDs, such that sources with spectra flatter than such limits are always "pure" AGN.  This means that far infrared SEDs can define a sample of AGN, but it will not be a complete sample because of the many AGN with cool components.  This is shown quantitatively in Figure 13 where the cool dust component as measured by $\nu$$f_{\nu}$(160 \ums)/$\nu$$f_{\nu}$(18 \ums) from AKARI fluxes is compared with strength of the starburst as measured by the PAH 6.2 \um feature. 

The values of this ratio show how starbursts have systematically cooler dust than AGN.  The median $\nu$$f_{\nu}$(160 \ums)/$\nu$$f_{\nu}$(18 \ums) for starbursts is 1.74, for composites is 1.71, and for AGN is 0.47.  These results indicate that the median "pure starburst" radiates almost 4 times more luminosity in the cool dust component relative to warm dust than does the median AGN.  The dispersion of points shows that sources which are AGN dominated in the mid-infrared, as determined by the strength of PAH, can nevertheless have cool dust components.  Pure starbursts cannot, however, have strong hot dust components, because the lower limit of this ratio for pure starbursts (0.7) is greater than the median for AGN (0.47).  

The resulting quantitative conclusion is that any source with $\nu$$f_{\nu}$(160 \ums)/$\nu$$f_{\nu}$(18 \ums) $<$ 0.7 must have part of the bolometric luminosity arising from an AGN.  Any source with $\nu$$f_{\nu}$(160 \ums)/$\nu$$f_{\nu}$(18 \ums) $<$ 0.5 can confidentally be classified as having dust continuum luminosity at all wavelengths arising primarily from an AGN, with negligible starburst contribution.


\subsection{Averages and Dispersions for Spectra}

 
The preceding results illustrate the dispersions in overall SEDs among the sources in the sample.  It is also useful to illustrate averages and dispersions within the IRS spectra.  Dispersions among spectra of different classifications are shown in Figure 14.  This result for the spectral dispersions is similar to the result discussed above for the photometric SEDs.   The largest dispersions are among the composite sources and absorbed AGN, noticeably larger than for pure starbursts.  

As emphasized in the Introduction, much of our motive for this study is to make comparisons between the local sources in this sample and the numerous sources at z $\ga$ 2 
which have been discovered with $Spitzer$ IRS spectroscopy.  To facilitate such comparisons, we show in Figure 15 the average spectra of sources with different classifications, as they would appear in the observed frame if redshifted to z = 2.5.  Only those portions of the spectrum with $\lambda$ $\la$ 35 \um are visible in IRS spectra; similarities in this wavelength range with the examples in Figure 1 are evident.

\section{Summary and Conclusions}

Spectroscopic results with the $Spitzer$ Infrared Spectrograph are compared to photometric SEDs to 160 \um from IRAS and AKARI for 301 sources.  Sources are selected to be spatially unresolved by comparing spectroscopic and photometric flux densities at 25 \ums, so that spectra and SEDs can be reliably compared.  Sources have 0.004 $<$ z $<$ 0.34 and 42.5 $<$ log $L_{IR}$ $<$ 46.8 (erg s$^{-1}$) and cover the full range of starburst galaxy and AGN classifications.  All spectra are newly determined with an optimal extraction procedure and made available electronically (http://cassis.astro.cornell.edu/sargsyan-etal). 

A consistent spectroscopic classification is used for sources and compared to optical classifications.  The infrared classification derives from EW of the 6.2 \um PAH emission, defining starbursts [EW(6.2 \ums) $>$ 0.4 \ums)], composite sources [0.1 \um $<$ EW(6.2 \ums) $<$ 0.4 \ums], and AGN [EW(6.2 \ums) $<$ 0.1 \ums].  

Total infrared luminosities $L_{IR}$ and SEDs are compared to the 7.7 \um PAH spectral peak for sources with PAH features or to the dust continuum at 7.9 \um for AGN without PAH features.  We find that log [$L_{IR}$/$\nu L_{\nu}$(7.7 \ums)] = 0.74 $\pm$ 0.18 in starbursts, that log [$L_{IR}$/$\nu L_{\nu}$(7.7 \ums)] = 0.96 $\pm$ 0.26 in composite sources (starburst plus AGN), that log [$L_{IR}$/$\nu L_{\nu}$(7.9 \ums)] = 0.80 $\pm$ 0.25 in AGN with silicate absorption, and log [$L_{IR}$/$\nu L_{\nu}$(7.9 \ums)] = 0.51 $\pm$ 0.21 in AGN with silicate emission.  There is no dependence of these ratios on $L_{IR}$.  

These scaling ratios are used to determine $L_{IR}$ from IRS spectra for the most luminous sources yet discovered (at z $\sim$ 2.5).  It is found that $L_{IR}$ are similar for the most luminous silicate absorption and silicate emission AGN, and the $L_{IR}$ for the most luminous AGN are about 2.5 times larger than for the most luminous starbursts.  

Starbursts show the most consistent spectra and SEDs at all wavelengths.  The dispersions generally range only over a factor of two at any wavelength.   This means that a mid-infrared classification as starburst is a good predictor of far infrared properties.   Sources with the strongest far infrared luminosity from cool dust components are composite sources, indicating that these sources may contain the most obscured starbursts.  Typical SEDs for AGN are flatter than for starbursts, but many objects classified as AGN in the mid-infrared spectrum have cool dust components as significant as in starbursts. 

Comparison of SEDs indicates that any source with $\nu$$f_{\nu}$(160 \ums)/$\nu$$f_{\nu}$(18 \ums) $<$ 0.7 must have part of the bolometric luminosity arising from an AGN.  Any source with $\nu$$f_{\nu}$(160 \ums)/$\nu$$f_{\nu}$(18 \ums) $<$ 0.5 can confidently be classified as having dust continuum luminosity at all wavelengths arising primarily from an AGN, with negligible starburst contribution.

\acknowledgments
This work is based primarily on observations made with the
Spitzer Space Telescope, which is operated by the Jet Propulsion
Laboratory, California Institute of Technology, under NASA contract
1407.  This work made use of the NASA/IPAC Extragalactic Data Base (NED) operated by JPL/Caltech under contract with NASA.  Support for this work by the IRS GTO team at Cornell University was provided by NASA through Contract
Number 1257184 issued by JPL/Caltech.

\clearpage

\begin{deluxetable}{lccccccccc} 

\tablecolumns{10}
\rotate
\tabletypesize{\scriptsize}
\tablewidth{0pc}
\tablecaption{Spectroscopic Properties of Starbursts}
\tablehead{
\colhead{source} & \colhead{Name} & \colhead{FSC Name} & \colhead{J2000 coordinates} & \colhead{z\tablenotemark{a}} & \colhead{SDSS\tablenotemark{b}}& \colhead{EW(6.2 \ums)\tablenotemark{c}} & \colhead{f(6.2 \ums)\tablenotemark{d}} & \colhead{f(11.3 \ums)\tablenotemark{e}}& \colhead{$f_{\nu}$(7.7 \ums)}  \\
\colhead{} & \colhead{} & \colhead{} & \colhead{} & \colhead{} & \colhead{} & \colhead{\ums} & \colhead {} & \colhead{} & \colhead{mJy}  }

\startdata
1 & Mrk938 & F00085-1223 & 001106.50-120626.0 & 0.0196 & \nodata & 0.44 & 33.50 & 22.10 & 475.0 \\%
2 & MCG-02-01-051 & F00163-1039 & 001850.90-102236.7 & 0.0271 & sb & 0.56 & 24.40 & 17.24 & 389.4 \\%
3 & IRAS00456-2904 & F00456-2904 & 004806.75-284818.6 & 0.1103 & \nodata & 0.43 & 3.06 & 2.26 & 56.9 \\%
4 & MCG+12-02-001 & F00506+7248 & 005403.88+730505.9 & 0.0157 & \nodata & 0.53 & 49.90 & 33.67 & 810.1 \\%
5 & MCG-03-04-014 & F01076-1707 & 011008.93-165109.9 & 0.0335 & \nodata & 0.45 & 25.26 & 24.39 & 513.8 \\%
6 & ESO244-G012 & F01159-4443 & 011808.31-442743.4 & 0.0229 & \nodata & 0.54 & 34.15 & 23.78 & 555.9 \\%
7 & ESO353-G020 & F01325-3623 & 013451.26-360814.4 & 0.0159 & \nodata & 0.40 & 31.51 & 23.40 & 634.0 \\%
8 & UGC01385 & F01519+3640 & 015453.82+365504.3 & 0.0188 & \nodata & 0.54 & 18.90 & 15.87 & 294.2 \\
\enddata
\tablenotetext{a}{Redshift from optical redshift in NASA/IPAC Extragalactic Data Base (NED), except source 18, from IRS spectrum. } 
\tablenotetext{b}{Our optical classification determined from spectra of the Sloan Digital Sky Survey, determined from presence of broad hydrogen emission lines or ratios among [OIII] $\lambda$5007, H$\beta$, [NII] $\lambda$6584, and H$\alpha$.}
\tablenotetext{c}{Rest frame equivalent width of 6.2 \um PAH feature fit with single gaussian on a linear continuum within rest wavelength range 5.5 \um to 6.9 \um.}
\tablenotetext{d}{Total flux of PAH 6.2 \um feature fit with single gaussian on a linear continuum within rest wavelength range 5.5 \um to 6.9 \um in units of 10$^{-20}$W cm$^{-2}$.}
\tablenotetext{e}{Total flux of PAH 11.3 \um feature fit with single gaussian on a linear continuum within rest wavelength range 10.5 \um to 12.0 \um in units of 10$^{-20}$W cm$^{-2}$.}

\end{deluxetable}

\clearpage

\begin{deluxetable}{lcccccccccccccccc} 

\tablecolumns{17}
\rotate
\tabletypesize{\scriptsize}
\tablewidth{0pc}
\tablecaption{SEDs and Total Luminosities of Starbursts}
\tablehead{
\colhead{source} & \colhead{FSC Name} & \colhead{  $f_{\nu}$(25 )} & \colhead{  $f_{\nu}$(8  )\tablenotemark{a}} & \colhead{  $f_{\nu}$(12  )\tablenotemark{b}} & \colhead{ $f_{\nu}$(18  )\tablenotemark{a}} & \colhead{ $f_{\nu}$(25  )\tablenotemark{b}} & \colhead{  $f_{\nu}$(60  )\tablenotemark{b}} & \colhead{ $f_{\nu}$(65  )\tablenotemark{a}}& \colhead{ $f_{\nu}$(90  )\tablenotemark{a}}& \colhead{  $f_{\nu}$(100 )\tablenotemark{b}}& \colhead{ $f_{\nu}$(140  )\tablenotemark{a}}& \colhead{ $f_{\nu}$(160  )\tablenotemark{a}} & \colhead{ $f_{\nu}$(1.25 )\tablenotemark{c}} & \colhead{ $f_{\nu}$(1.66 )\tablenotemark{c}}  & \colhead{R\tablenotemark{d}} & \colhead{$L_{ir}$\tablenotemark{e}}\\
\colhead{} & \colhead{} & \colhead{IRS Jy} & \colhead{Jy} & \colhead{Jy} & \colhead{Jy} & \colhead{Jy} & \colhead{Jy} & \colhead{Jy}  & \colhead{Jy} & \colhead{Jy} & \colhead{Jy} & \colhead{Jy} &  \colhead{mJy} &  \colhead{mJy}  &  \colhead{} & \colhead{log}  }

\startdata
1 & F00085-1223 & 1.563 & 0.269 & 0.350 & \nodata & 2.390 & 17.050 & 14.430 & 17.041 & 16.860 & 12.127 & 11.149 & 18.97 & 29.61 & 7.73 & 45.07 \\%
2 & F00163-1039 & 0.997 & 0.178 & 0.270 & 0.517 & 1.130 & 6.880 & 5.918 & 8.251 & 9.620 & 7.303 & 6.951 & 4.49 & 6.38 & 4.49 & 45.03 \\%
3 & F00456-2904 & 0.161 & \nodata & 0.084 & \nodata & 0.141 & 2.600 & 2.557 & 2.479 & 3.380 & 4.084 & \nodata & 1.41 & 1.78 & 10.77 & 45.79 \\%
4 & F00506+7248 & 2.561 & 0.493 & 0.765 & 1.668 & 3.720 & 22.400 & 15.707 & 26.171 & 26.700 & 19.092 & 19.151 & 7.73 & 11.61 & 6.60 & 45.04 \\%
5 & F01076-1707 & 0.737 & 0.203 & 0.301 & 0.511 & 0.846 & 6.480 & 5.944 & 8.097 & 10.400 & 10.367 & 5.217 & 8.52 & 14.98 & 3.30 & 45.20 \\%
6 & F01159-4443 & 1.585 & 0.256 & 0.303 & 0.805 & 1.860 & 7.840 & 7.184 & 10.033 & 11.700 & 8.497 & 6.698 & 12.38 & 17.60 & 3.88 & 44.97 \\%
7 & F01325-3623 & 0.550 & 0.226 & 0.258 & 0.341 & 0.677 & 7.290 & 7.159 & 11.084 & 15.800 & 16.619 & 13.424 & 10.3 & 17.73 & 3.08 & 44.61 \\%
8 & F01519+3640 & 0.861 & 0.151 & 0.211 & 0.430 & 0.962 & 5.490 & 5.118 & 5.705 & 7.780 & 6.296 & 4.514 & 13.16 & 17.88 & 4.76 & 44.61 \\
\enddata
\tablenotetext{a}{Flux densities at listed wavelengths taken from AKARI catalog, http://darts.isas.jaxa.jp/astro/akari/cas/tools/search/crossid.html.}
\tablenotetext{b}{Flux densities at listed wavelengths taken from IRAS Faint Source Catalog, http://vizier.u-strasbg.fr/viz-bin/VizieR-4.}
\tablenotetext{c}{Flux densities at listed wavelengths taken from 2MASS catalog.}
\tablenotetext{d}{Ratio R = $f_{IR}$/$\nu$$f_{\nu}$(7.7 \ums), for $f_{IR}$ =  1.8 x 10$^{-11}$[13.48$f_{\nu}$(12) + 5.16$f_{\nu}$(25) + 2.58$f_{\nu}$(60) + $f_{\nu}$(100)] in erg cm$^{-2}$ s$^{-1}$.}
\tablenotetext{e}{Total infrared luminosity $L_{ir}$ =  4$\pi$D$^{2}$$f_{IR}$ in erg s$^{-1}$ for D the luminosity distance determined for H$_0$ = 71 \kmsMpc, $\Omega_{M}$=0.27 and $\Omega_{\Lambda}$=0.73, from \citet{wri06}:  http://www.astro.ucla.edu/~wright/CosmoCalc.html. }

\end{deluxetable}

\clearpage

\begin{deluxetable}{lccccccccc} 

\tablecolumns{10}
\rotate
\tabletypesize{\scriptsize}
\tablewidth{0pc}
\tablecaption{Spectroscopic Properties of Composite Sources}
\tablehead{
\colhead{source} & \colhead{Name} & \colhead{FSC Name} & \colhead{J2000 coordinates} & \colhead{z\tablenotemark{a}} & \colhead{SDSS\tablenotemark{b}}& \colhead{EW(6.2 \ums)\tablenotemark{c}} & \colhead{f(6.2 \ums)\tablenotemark{d}} & \colhead{f(11.3 \ums)\tablenotemark{e}}& \colhead{$f_{\nu}$(7.7 \ums)}  \\
\colhead{} & \colhead{} & \colhead{} & \colhead{} & \colhead{} & \colhead{} & \colhead{\ums} & \colhead {} & \colhead{} & \colhead{mJy}  }

\startdata
1 & Mrk334 & F00006+2141 & 000309.62+215736.6 & 0.0219 & \nodata & 0.25 & 17.31 & 12.15 & 334.6 \\%
2 & IRAS00199-7426 & F00199-7426 & 002207.01-740941.7 & 0.0964 & \nodata & 0.35 & 5.54 & 3.83 & 113.2 \\%
3 & E12-G21 & F00392-7930 & 004046.10-791424.0 & 0.0300 & \nodata & 0.28 & 10.90 & 7.53 & 168.8 \\%
4 & NGC0291 & F00509-0902 & 005329.92-084604.1 & 0.0190 & agn & 0.33 & 5.17 & 4.97 & 111.4 \\%
5 & CGCG436-030 & F01173+1405 & 012002.63+142142.3 & 0.0312 & sb & 0.23 & 15.06 & 9.15 & 361.0 \\%
6 & IRASF01197+0044 & F01197+0044 & 012218.13+010026.0 & 0.0555 & sb & 0.14 & 3.23 & 1.86 & 83.7 \\%
7 & IRASF01364-1042 & F01364-1042 & 013852.91-102711.0 & 0.0482 & agn & 0.31 & 2.82 & 1.48 & 66.6 \\%
8 & IIIZw035 & F01417+1651 & 014430.56+170609.0 & 0.0274 & \nodata & 0.37 & 4.48 & 4.93 & 106.8 \\
\enddata
\tablenotetext{a}{Redshift from optical redshift in NED. } 
\tablenotetext{b}{Our optical classification determined from spectra of the Sloan Digital Sky Survey, determined from presence of broad hydrogen emission lines or ratios among [OIII] $\lambda$5007, H$\beta$, [NII] $\lambda$6584, and H$\alpha$.}
\tablenotetext{c}{Rest frame equivalent width of 6.2 \um PAH feature fit with single gaussian on a linear continuum within rest wavelength range 5.5 \um to 6.9 \um.}
\tablenotetext{d}{Total flux of PAH 6.2 \um feature fit with single gaussian on a linear continuum within rest wavelength range 5.5 \um to 6.9 \um in units of 10$^{-20}$W cm$^{-2}$.}
\tablenotetext{e}{Total flux of PAH 11.3 \um feature fit with single gaussian on a linear continuum within rest wavelength range 10.5 \um to 12.0 \um in units of 10$^{-20}$W cm$^{-2}$.}
\tablenotetext{f}{$f_{\nu}$(7.9 \ums) measured because continuum stronger than PAH peak $f_{\nu}$(7.7 \ums).}

\end{deluxetable}

\clearpage

\begin{deluxetable}{lcccccccccccccccc} 

\tablecolumns{17}
\rotate
\tabletypesize{\scriptsize}
\tablewidth{0pc}
\tablecaption{SEDs and Total Luminosities of Composite Sources}
\tablehead{
\colhead{source} & \colhead{FSC Name} & \colhead{  $f_{\nu}$(25 )} & \colhead{  $f_{\nu}$(8  )\tablenotemark{a}} & \colhead{  $f_{\nu}$(12  )\tablenotemark{b}} & \colhead{ $f_{\nu}$(18  )\tablenotemark{a}} & \colhead{ $f_{\nu}$(25  )\tablenotemark{b}} & \colhead{  $f_{\nu}$(60  )\tablenotemark{b}} & \colhead{ $f_{\nu}$(65  )\tablenotemark{a}}& \colhead{ $f_{\nu}$(90  )\tablenotemark{a}}& \colhead{  $f_{\nu}$(100 )\tablenotemark{b}}& \colhead{ $f_{\nu}$(140  )\tablenotemark{a}}& \colhead{ $f_{\nu}$(160  )\tablenotemark{a}} & \colhead{ $f_{\nu}$(1.25 )\tablenotemark{c}} & \colhead{ $f_{\nu}$(1.66 )\tablenotemark{c}}  & \colhead{R\tablenotemark{d}} & \colhead{$L_{ir}$\tablenotemark{e}}\\
\colhead{} & \colhead{} & \colhead{IRS Jy} & \colhead{Jy} & \colhead{Jy} & \colhead{Jy} & \colhead{Jy} & \colhead{Jy} & \colhead{Jy}  & \colhead{Jy} & \colhead{Jy} & \colhead{Jy} & \colhead{Jy} &  \colhead{mJy} &  \colhead{mJy}  &  \colhead{} & \colhead{log}  }

\startdata
1 & F00006+2141 & 0.959 & 0.197 & 0.226 & 0.546 & 1.050 & 4.340 & 3.212 & 3.715 & 4.320 & 4.340 & 2.109 & 10.91 & 17.39 & 3.38 & 44.66 \\%
2 & F00199-7426 & 0.265 & \nodata & 0.105 & \nodata & 0.326 & 4.160 & 3.948 & 4.403 & 6.420 & 5.416 & 4.198 & 2.84 & 4.09 & 9.06 & 45.90 \\%
3 & F00392-7930 & 0.194 & 0.139 & 0.220 & 0.182 & 0.190 & 1.510 & 1.412 & 1.632 & 3.220 & 4.007 & 2.986 & 11 & 17.29 & 3.12 & 44.60 \\%
4 & F00509-0902 & 0.385 & \nodata & 0.313 & 0.204 & 0.529 & 2.610 & 2.631 & 3.135 & 3.800 & 3.745 & 2.577 & 5.6 & 6.98 & 7.39 & 44.39 \\%
5 & F01173+1405 & 1.333 & 0.182 & 0.265 & 0.592 & 1.440 & 11.100 & 10.669 & 10.441 & 9.330 & 8.062 & 4.886 & 4.55 & 11.28 & 6.46 & 45.28 \\%
6 & F01197+0044 & 0.275 & \nodata & 0.110 & 0.306 & 0.378 & 2.700 & 2.258 & 2.320 & 4.450 & 2.971 & 2.953 & 0.92 & 0.90 & 8.65 & 45.27 \\%
7 & F01364-1042 & 0.335 & \nodata & 0.077 & \nodata & 0.396 & 6.160 & 5.018 & 6.289 & 6.700 & 4.802 & 1.669 & 1.42 & 2.05 & 18.67 & 45.38 \\%
8 & F01417+1651 & 0.936 & \nodata & 0.139 & 0.324 & 0.947 & 12.600 & 11.711 & 13.123 & 13.300 & 7.740 & 6.470 & 6.46 & 8.71 & 23.36 & 45.19 \\
\enddata
\tablenotetext{a}{Flux densities at listed wavelengths taken from AKARI catalog, http://darts.isas.jaxa.jp/astro/akari/cas/tools/search/crossid.html.}
\tablenotetext{b}{Flux densities at listed wavelengths taken from IRAS Faint Source Catalog, http://vizier.u-strasbg.fr/viz-bin/VizieR-4.}
\tablenotetext{c}{Flux densities at listed wavelengths taken from 2MASS catalog.}
\tablenotetext{d}{Ratio R = $f_{IR}$/$\nu$$f_{\nu}$(7.7 \ums), for $f_{IR}$ =  1.8 x 10$^{-11}$[13.48$f_{\nu}$(12) + 5.16$f_{\nu}$(25) + 2.58$f_{\nu}$(60) + $f_{\nu}$(100)] in erg cm$^{-2}$ s$^{-1}$.}
\tablenotetext{e}{Total infrared luminosity $L_{ir}$ =  4$\pi$D$^{2}$$f_{IR}$ in erg s$^{-1}$ for D the luminosity distance determined for H$_0$ = 71 \kmsMpc, $\Omega_{M}$=0.27 and $\Omega_{\Lambda}$=0.73, from \citet{wri06}:  http://www.astro.ucla.edu/~wright/CosmoCalc.html. }
\tablenotetext{f}{$f_{\nu}$(7.9\ums) measured for ratio because continuum stronger than PAH peak $f_{\nu}$(7.7 \ums).}

\end{deluxetable}

\clearpage

\begin{deluxetable}{lcccccccccc} 

\tablecolumns{11}
\rotate
\tabletypesize{\scriptsize}
\tablewidth{0pc}
\tablecaption{Spectroscopic Properties of AGN}
\tablehead{
\colhead{source} & \colhead{Name} & \colhead{FSC Name} & \colhead{J2000 coordinates} & \colhead{z\tablenotemark{a}} & \colhead{SDSS\tablenotemark{b}}& \colhead{EW(6.2 \ums)\tablenotemark{c}} & \colhead{f(6.2 \ums)\tablenotemark{d}} & \colhead{f(11.3 \ums)\tablenotemark{e}}& \colhead{$f_{\nu}$(7.7 \ums)} & \colhead{silicate\tablenotemark{f}}  \\
\colhead{} & \colhead{} & \colhead{} & \colhead{} & \colhead{} & \colhead{} & \colhead{\ums} & \colhead {} & \colhead{} & \colhead{mJy} & \colhead{}  }

\startdata
1 & IRAS00275-2859 & F00275-2859 & 003004.20-284225.4 & 0.2781 & \nodata & 0.011 & 0.22 & 0.19 & 63.4\tablenotemark{g} & 0.71 \\%
2 & IRAS00406-3127 & F00406-3127 & 004303.14-311049.7 & 0.3424 & \nodata & $<$0.011 & $<$0.08 & 0.24 & 34.5\tablenotemark{g} & 0.15 \\%
3 & IZw1 & F00509+1225 & 005334.94+124136.2 & 0.0611 & \nodata & $<$0.003 & $<$0.48 & 1.25 & 294.6\tablenotemark{g} & 1.27 \\%
4 & IRAS01003-2238 & F01004-2237 & 010249.94-222157.3 & 0.1177 & \nodata & $<$0.043 & $<$0.99 & 0.28 & 106.5\tablenotemark{g} & 0.47 \\%
5 & NGC454 & F01123-5539 & 011424.90-552352.0 & 0.0122 & \nodata & $<$0.008 & $<$0.66 & 3.16 & 130.5 & P+abs \\%
6 & IRAS01199-2307 & F01199-2307 & 012220.84-225157.3 & 0.1560 & \nodata & $<$0.088 & $<$0.41 & 0.22 & 24.1\tablenotemark{g} & 0.03 \\%
7 & NGC526a & F01216-3519 & 012354.39-350355.0 & 0.0191 & \nodata & $<$0.001 & $<$0.08 & $<$0.08 & 112.9 & \nodata \\%
8 & 3C48 & F01348+3254 & 013741.29+330935.0 & 0.3670 & \nodata & 0.006 & 0.31 & 0.25 & 60.5 & \nodata \\
\enddata
\tablenotetext{a}{Redshift from optical redshift in NED unless noted.} 
\tablenotetext{b}{Our optical classification determined from spectra of the Sloan Digital Sky Survey, determined from presence of broad hydrogen emission lines or ratios among [OIII] $\lambda$5007, H$\beta$, [NII] $\lambda$6584, and H$\alpha$.}
\tablenotetext{c}{Rest frame equivalent width of 6.2 \um PAH feature fit with single gaussian on a linear continuum within rest wavelength range 5.5 \um to 6.9 \um.}
\tablenotetext{d}{Total flux of PAH 6.2 \um feature fit with single gaussian on a linear continuum within rest wavelength range 5.5 \um to 6.9 \um in units of 10$^{-20}$W cm$^{-2}$.}
\tablenotetext{e}{Total flux of PAH 11.3 \um feature fit with single gaussian on a linear continuum within rest wavelength range 10.5 \um to 12 \um in units of 10$^{-20}$W cm$^{-2}$.} 
\tablenotetext{f} {Strength of silicate feature defined as $f_{\nu}$(10 \um observed)/$f_{\nu}$(10 \um continuum), for $f_{\nu}$(10 \um continuum) extrapolated linearly between $f_{\nu}$(7.9 \ums) and $f_{\nu}$(13 \ums).  Values $>$ 1 correspond to silicate emission and values $<$ 1 to silicate absorption. Entry P+abs means that both PAH emission and silicate absorption are present; absorption strength is not measured because continuum level at 7.9 \um cannot be reliably determined.}
\tablenotetext{g}{$f_{\nu}$(7.9 \ums) measured because continuum peak stronger than $f_{\nu}$(7.7 \ums) at PAH feature.}
\tablenotetext{h}{Redshift from IRS spectrum.}
\tablenotetext{i}{Unique object having optical classification like Blue Compact Dwarf (BCD) with narrow optical emission lines, strong [OIII]/H$\beta$ and very strong H$\alpha$/[NII].}

\end{deluxetable}

\clearpage

\begin{deluxetable}{lcccccccccccccccc} 

\tablecolumns{17}
\rotate
\tabletypesize{\scriptsize}
\tablewidth{0pc}
\tablecaption{SEDs and Total Luminosities of AGN}
\tablehead{
\colhead{source} & \colhead{FSC Name} & \colhead{  $f_{\nu}$(25 )} & \colhead{  $f_{\nu}$(8  )\tablenotemark{a}} & \colhead{  $f_{\nu}$(12  )\tablenotemark{b}} & \colhead{ $f_{\nu}$(18  )\tablenotemark{a}} & \colhead{ $f_{\nu}$(25  )\tablenotemark{b}} & \colhead{  $f_{\nu}$(60  )\tablenotemark{b}} & \colhead{ $f_{\nu}$(65  )\tablenotemark{a}}& \colhead{ $f_{\nu}$(90  )\tablenotemark{a}}& \colhead{  $f_{\nu}$(100 )\tablenotemark{b}}& \colhead{ $f_{\nu}$(140  )\tablenotemark{a}}& \colhead{ $f_{\nu}$(160  )\tablenotemark{a}} & \colhead{ $f_{\nu}$(1.25 )\tablenotemark{c}} & \colhead{ $f_{\nu}$(1.66 )\tablenotemark{c}}  & \colhead{R\tablenotemark{d}} & \colhead{$L_{ir}$\tablenotemark{e}}\\
\colhead{} & \colhead{} & \colhead{IRS Jy} & \colhead{Jy} & \colhead{Jy} & \colhead{Jy} & \colhead{Jy} & \colhead{Jy} & \colhead{Jy}  & \colhead{Jy} & \colhead{Jy} & \colhead{Jy} & \colhead{Jy} &  \colhead{mJy} &  \colhead{mJy}  &  \colhead{} & \colhead{log}  }

\startdata
1 & F00275-2859 & 0.144 & \nodata & 0.081 & 0.137 & 0.173 & 0.690 & 0.352 & 0.650 & 0.734 & \nodata & 1.569 & 1.31 & 2.69 & 4.3\tablenotemark{f} & 46.24 \\%
2 & F00406-3127 & 0.073 & \nodata & 0.060 & \nodata & 0.091 & 0.717 & \nodata & \nodata & 0.994 & \nodata & \nodata & \nodata & \nodata & 7.6\tablenotemark{f} & 46.40 \\%
3 & F00509+1225 & 0.989 & 0.321 & 0.512 & 0.714 & 1.210 & 2.240 & 1.606 & 1.964 & 2.630 & 2.575 & 3.055 & 16.63 & 25.98 & 3.68\tablenotemark{f} & 45.52 \\%
4 & F01004-2237 & 0.538 & \nodata & 0.227 & 0.402 & 0.660 & 2.290 & 3.063 & 1.376 & 1.790 & 1.537 & \nodata & 0.59 & 0.80 & 7.05\tablenotemark{f} & 45.93 \\%
5 & F01123-5539 & 0.333 & \nodata & 0.213 & \nodata & 0.417 & 1.480 & \nodata & \nodata & 2.820 & \nodata & \nodata & 20.37 & 28.05 & 4.18 & 43.83 \\%
6 & F01199-2307 & 0.111 & \nodata & 0.106 & \nodata & 0.162 & 1.610 & 1.698 & 1.002 & 1.370 & 1.298 & 4.932 & \nodata & \nodata & 17.71\tablenotemark{f} & 45.93 \\%
7 & F01216-3519 & 0.347 & 0.141 & 0.285 & 0.292 & 0.449 & 0.224 & \nodata & \nodata & 0.872 & \nodata & \nodata & 12.08 & 17.78 & 3.17 & 44.03 \\%
8 & F01348+3254 & 0.155 & \nodata & 0.068 & \nodata & 0.161 & 0.740 & 1.164 & 0.695 & 1.150 & 0.257 & 1.568 & 2.34 & 3.08 & 5.02 & 46.54 \\
\enddata
\tablenotetext{a}{Flux densities at listed wavelengths taken from AKARI catalog, http://darts.isas.jaxa.jp/astro/akari/cas/tools/search/crossid.html.}
\tablenotetext{b}{Flux densities at listed wavelengths taken from IRAS Faint Source Catalog, http://vizier.u-strasbg.fr/viz-bin/VizieR-4.}
\tablenotetext{c}{Flux densities at listed wavelengths taken from 2MASS catalog.}
\tablenotetext{d}{Ratio R = $f_{IR}$/$\nu$$f_{\nu}$(7.7 or 7.9 \ums), for $f_{IR}$ =  1.8 x 10$^{-11}$[13.48$f_{\nu}$(12) + 5.16$f_{\nu}$(25) + 2.58$f_{\nu}$(60) + $f_{\nu}$(100)] in erg cm$^{-2}$ s$^{-1}$.}
\tablenotetext{e}{Total infrared luminosity $L_{ir}$ =  4$\pi$D$^{2}$$f_{IR}$ in erg s$^{-1}$ for D the luminosity distance determined for H$_0$ = 71 \kmsMpc, $\Omega_{M}$=0.27 and $\Omega_{\Lambda}$=0.73, from \citet{wri06}:  http://www.astro.ucla.edu/~wright/CosmoCalc.html. }
\tablenotetext{f}{$f_{\nu}$(7.9 \ums) measured because continuum peak stronger than $f_{\nu}$(7.7 \ums) at PAH feature.}
\tablenotetext{g}{Unique object, having highest value of R accompanied by unusually low $L_{ir}$; no SDSS spectrum but optical image resembles BCD.}
\tablenotetext{h}{Unique object noted in Table 5 with weak PAH and optical spectrum like low luminosity BCD \citep [e.g.][]{hun10}, but $L_{ir}$ compares to normal starbursts and AGN.}

\end{deluxetable}









\begin{figure}
\figurenum{1}
\includegraphics[scale=0.9]{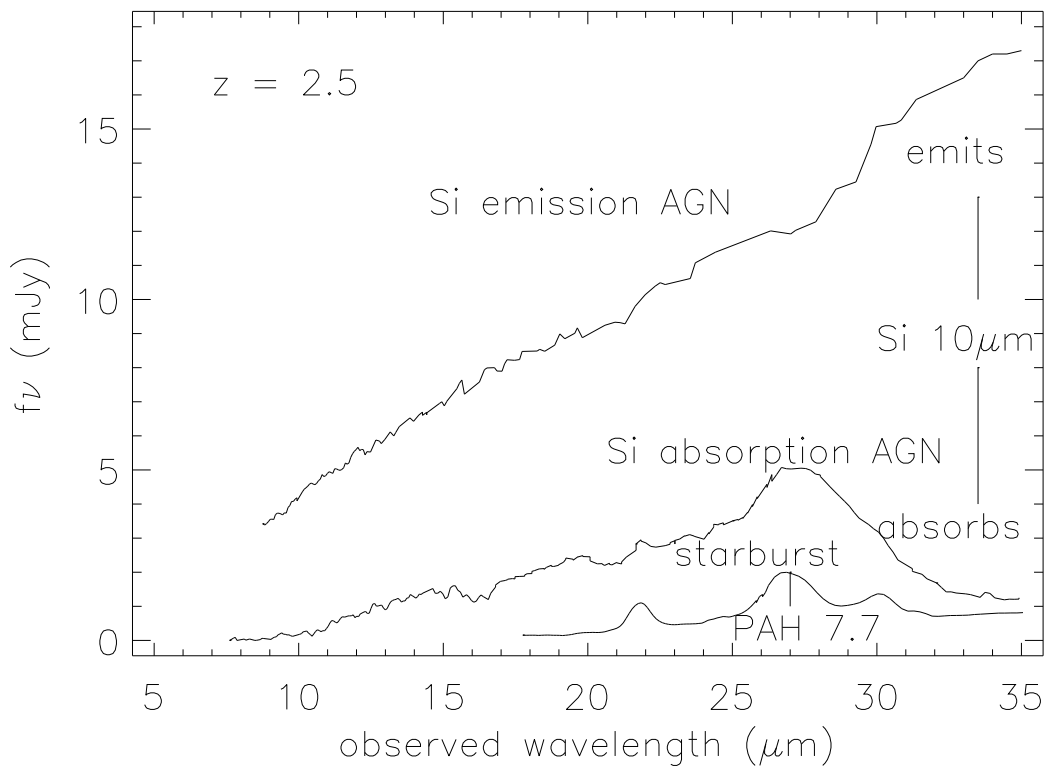}
\caption{Representative spectra of AGN and starbursts as would be observed at z = 2.5.  Spectra indicate why the parameter $\nu$$f_{\nu}$($\sim$8 \ums) (rest frame) is used because it is a straightforward measurement of the spectral peak in high redshift sources with poor S/N.  Flux densities represent the brightest sources known in each category at this redshift. } 

\end{figure}

\begin{figure}
\figurenum{2}
\includegraphics[scale=0.9]{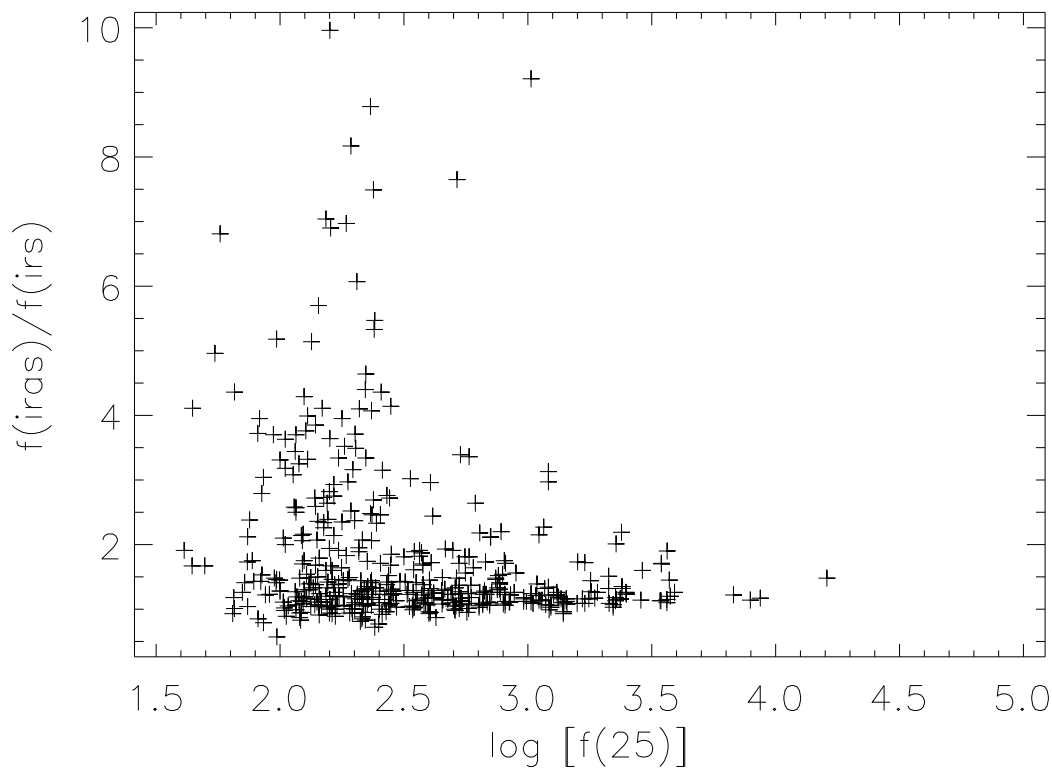}
\caption{Ratio of actual $f_{\nu}$(25 \ums) observed with IRAS, f(IRAS), to synthetic $f_{\nu}$(25 \ums) measured on IRS spectrum, f(IRS), compared to IRAS $f_{\nu}$(25 \ums) in mJy.  If ratio $\sim$ 1, it means that source is unresolved with the 10 \arcsec~ IRS LL1 slit so that IRAS and IRS measure the same source.   Larger ratios correspond to more extended sources.  Using these results, we study in this paper only those sources for which f(IRAS)/f(IRS) $<$ 1.5.  } 

\end{figure}

\begin{figure}
\figurenum{3}
\includegraphics[scale=0.9]{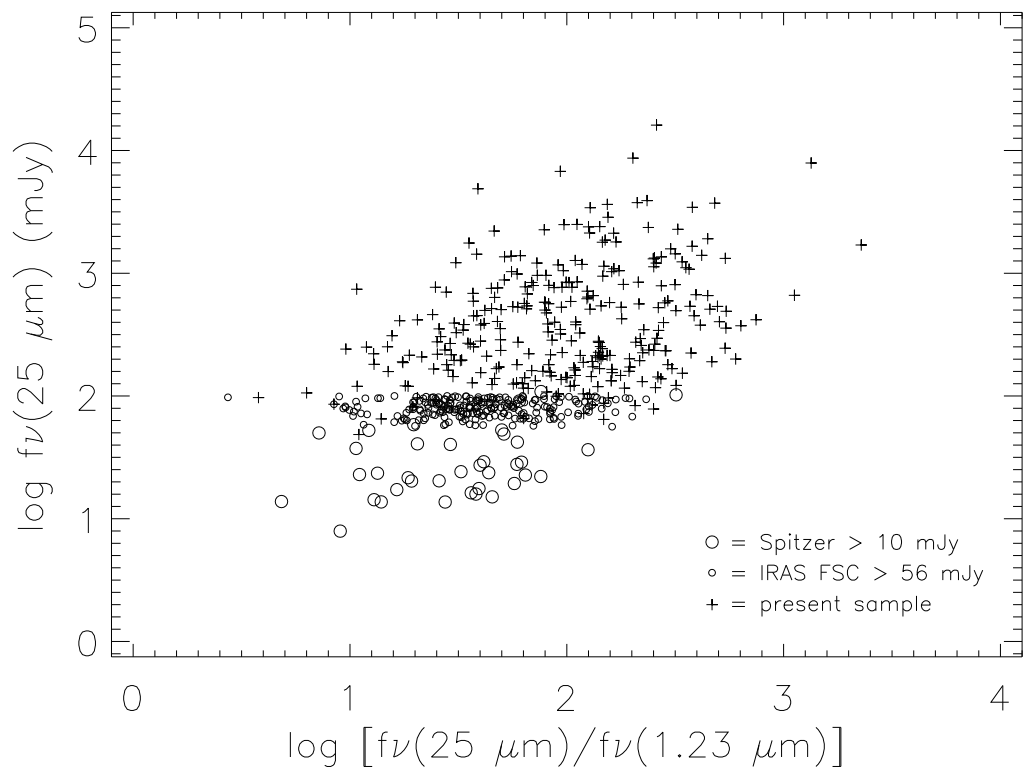}
\caption{ Range of ratios of mid-infrared (from dust) measured as IRAS $f_{\nu}$(25 \ums) to near-infrared (from stars or non-thermal AGN continuum) measured as 2MASS $J$ band flux density.  Crosses are the sources from our present sample in Tables 1-6; small circles are all IRAS Faint Source Catalog sources with 56 mJy $<$ $f_{\nu}$(25 \ums) $<$ 100 mJy, the faintest FSC sample \citep{hov10};  large circles are the $Spitzer$ flux limited sample with $f_{\nu}$(24 \ums) $>$ 10 mJy from \citet{wee09a}.  } 

\end{figure}

\begin{figure}
\figurenum{4}
\includegraphics[scale=0.9]{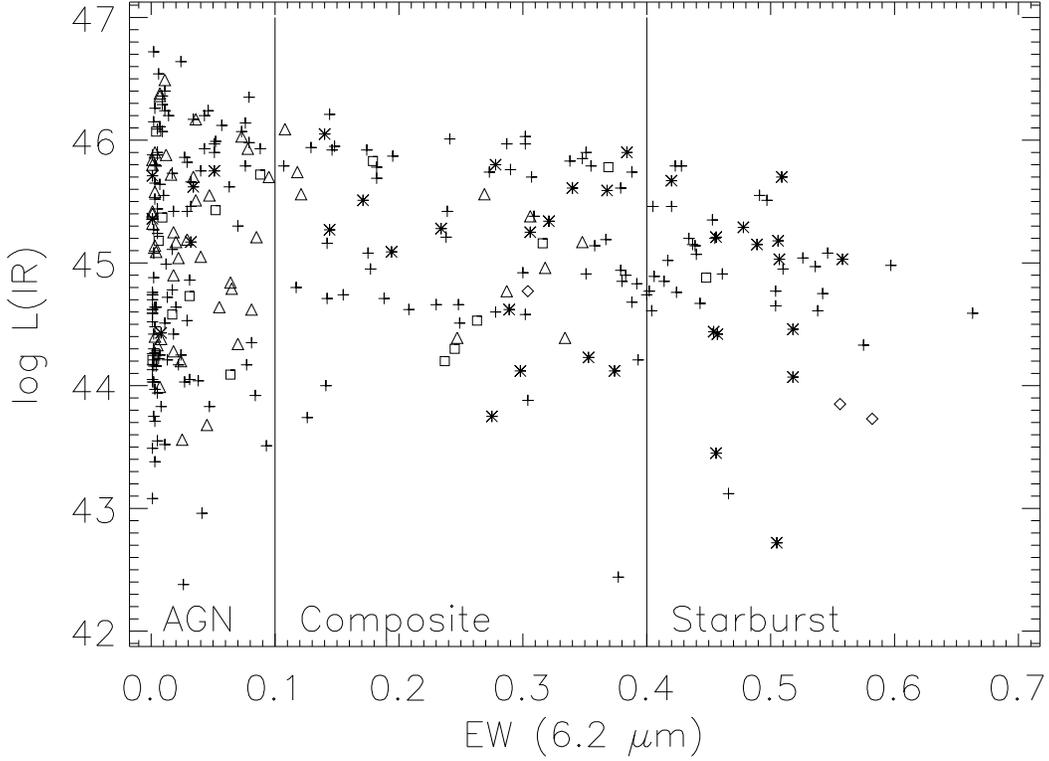}
\caption{Infrared luminosity $L_{IR}$ in erg s$^{-1}$ measured from total IRAS flux compared to equivalent width of 6.2 \um PAH feature, in \ums.  Symbols indicate optical classification if available from SDSS spectra, determined from presence of broad hydrogen emission lines or ratios among [OIII] $\lambda$5007, H$\beta$, [NII] $\lambda$6584, and H$\alpha$; triangles are optically classified AGN, squares are optically classified composite sources, asterisks are optically classified starbursts, and diamonds are absorption line sources not classifiable with emission lines; crosses are sources without SDSS spectra available for optical classification. Vertical lines divide sources classified according to infrared classification from EW(6.2 \ums), with "pure" AGN defined by EW(6.2 \ums) $<$ 0.1 \ums, "pure" starbursts, defined as EW(6.2 \ums) $>$ 0.4 \ums, and composite starburst+AGN defined as 0.1 \um $<$ EW(6.2 \ums) $<$ 0.4 \um.}
\end{figure}

\begin{figure}
\figurenum{5}
\includegraphics[scale=0.9]{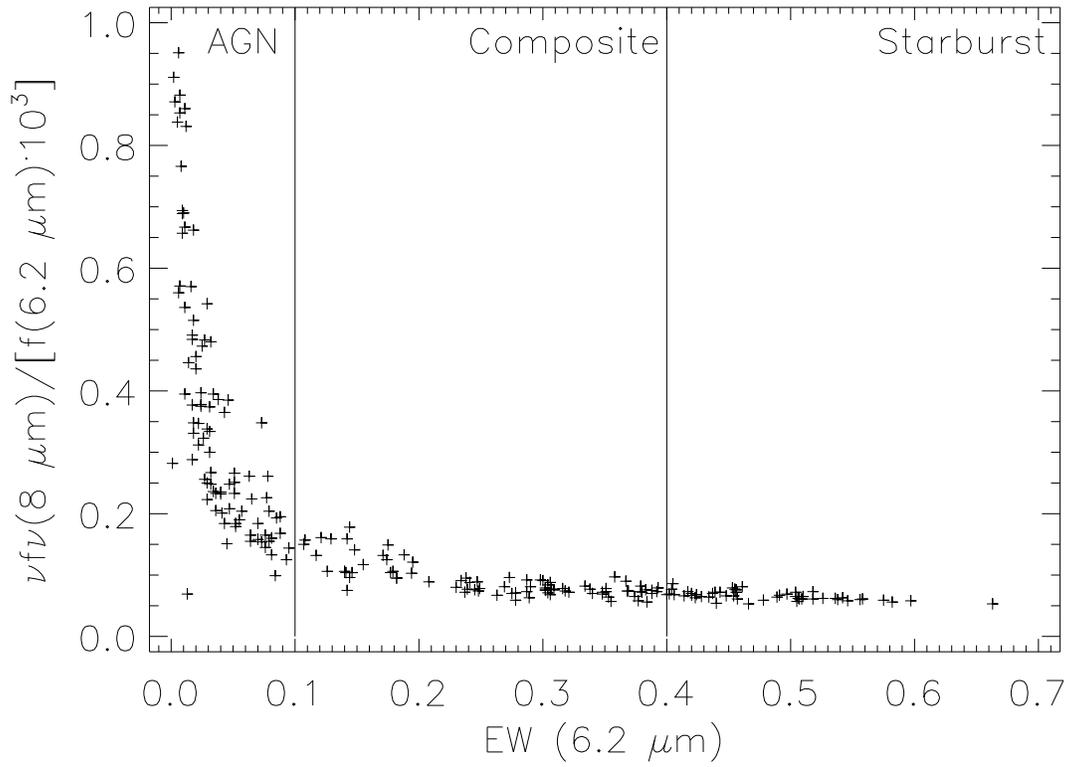}
\caption{Comparison of peak flux $\nu$$f_{\nu}$($\sim$ 8 \ums) with total flux measured in PAH 6.2 \um feature.  The peak flux is the peak of the 7.7 \um PAH feature if this feature is the local spectral maximum; for AGN without PAH features, the peak is the continuum at 7.9 \um.   }

\end{figure}

\begin{figure}
\figurenum{6}
\includegraphics[scale=0.9]{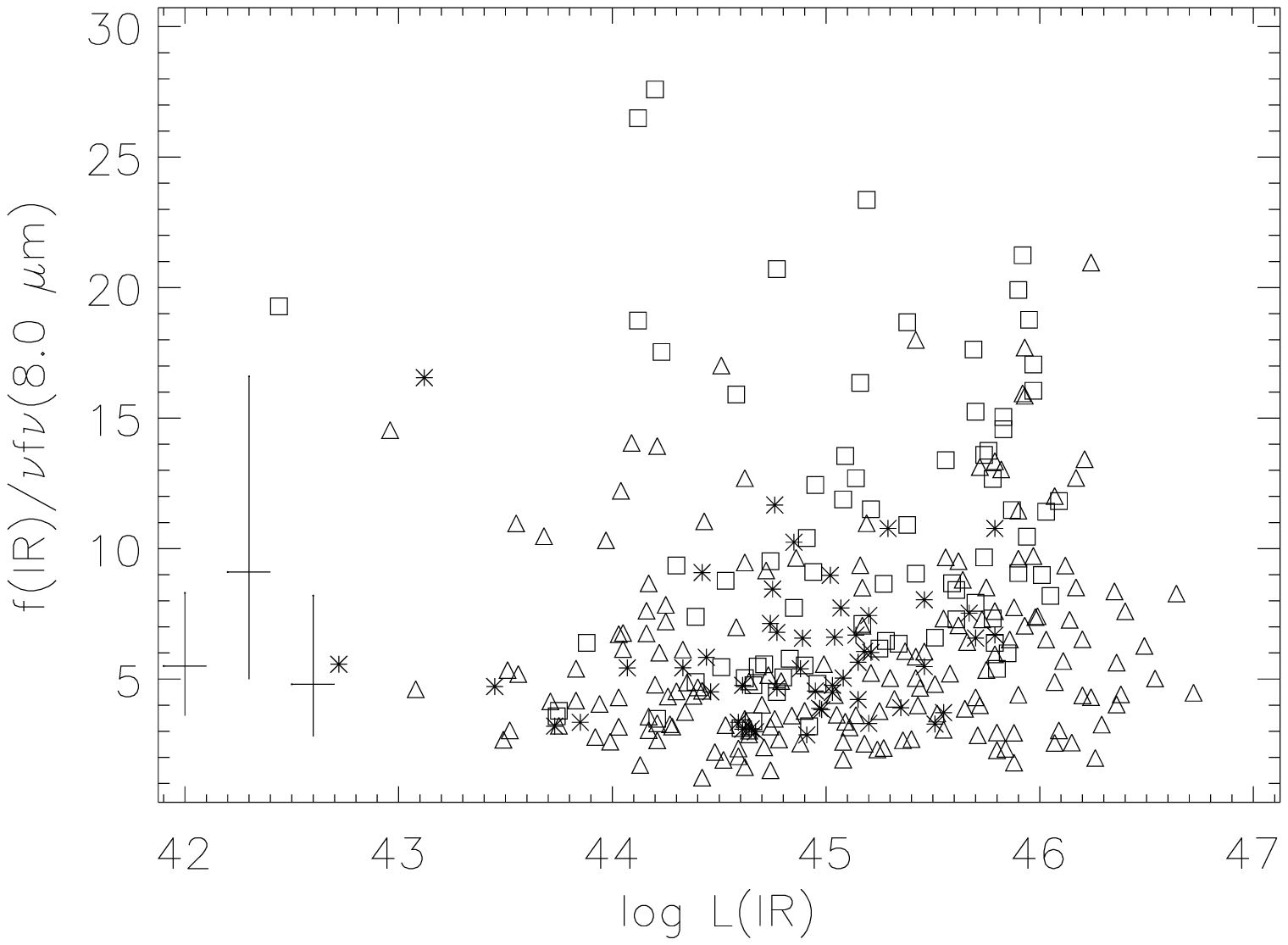}
\caption{Ratio $f_{IR}$/$\nu$$f_{\nu}$($\sim$ 8 \ums) for all sources compared to infrared luminosity $L_{IR}$ in erg s$^{-1}$ measured from total IRAS flux.   Asterisks are starbursts, defined as sources with EW(6.2 \ums) $>$ 0.4 \um for which spectral peak is the 7.7 \um PAH feature; squares are composite sources with 0.1 \um $<$ EW(6.2 \ums) $<$ 0.4 \um for which spectral peak is usually the 7.7 \um PAH feature except for sources noted in Table 3 for which continuum peak at 7.9 \um is stronger than PAH peak; triangles are AGN for which spectral peak is either the 7.7 \um PAH feature or the 7.9 \um continuum as noted in Table 5.  Error bars show medians and dispersions for starbursts, composites, and AGN (left to right). }

\end{figure}


\begin{figure}
\figurenum{7}
\includegraphics[scale=0.9]{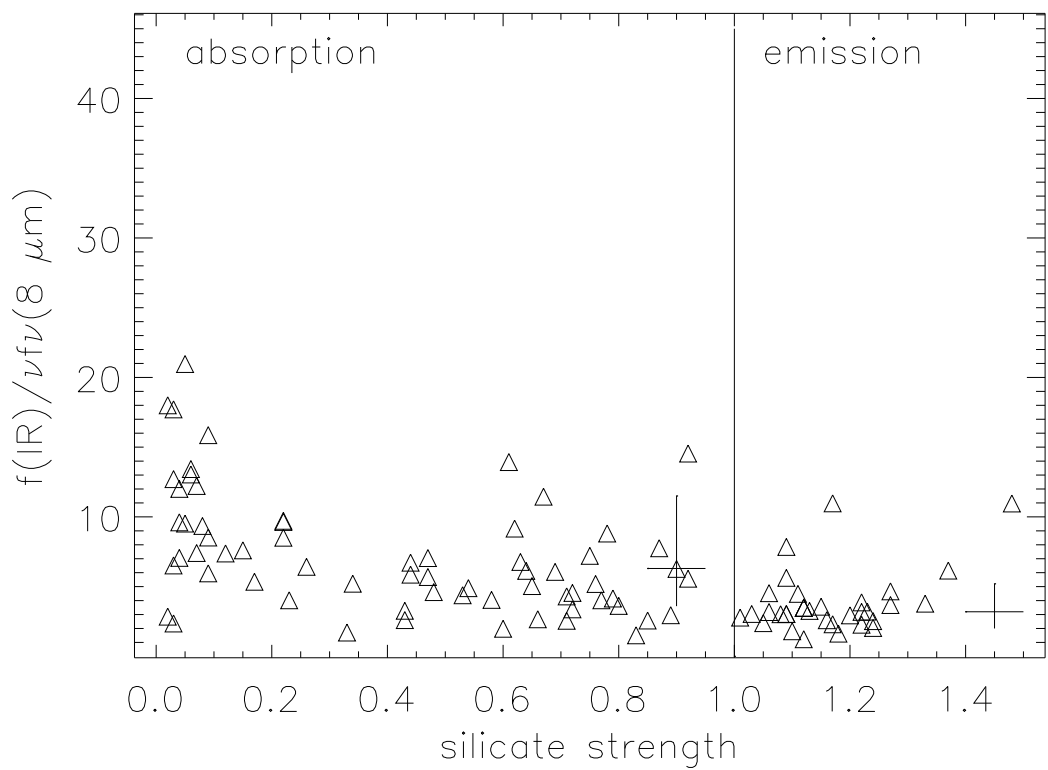}
\caption{Ratio $f_{IR}$/$\nu$$f_{\nu}$($\sim$8 \ums) for AGN sources compared to strength of silicate feature.  Strength of feature is defined as $f_{\nu}$(10 \um observed)/$f_{\nu}$(10 \um continuum), for $f_{\nu}$(10 \um continuum) extrapolated linearly between $f_{\nu}$(7.9 \ums) and $f_{\nu}$(13 \ums).  Values $>$ 1 correspond to silicate emission and values $<$ 1 to silicate absorption.  Error bars show medians and dispersions for absorption and emission AGN. }

\end{figure}

\begin{figure}
\figurenum{8}
\includegraphics[scale=0.9]{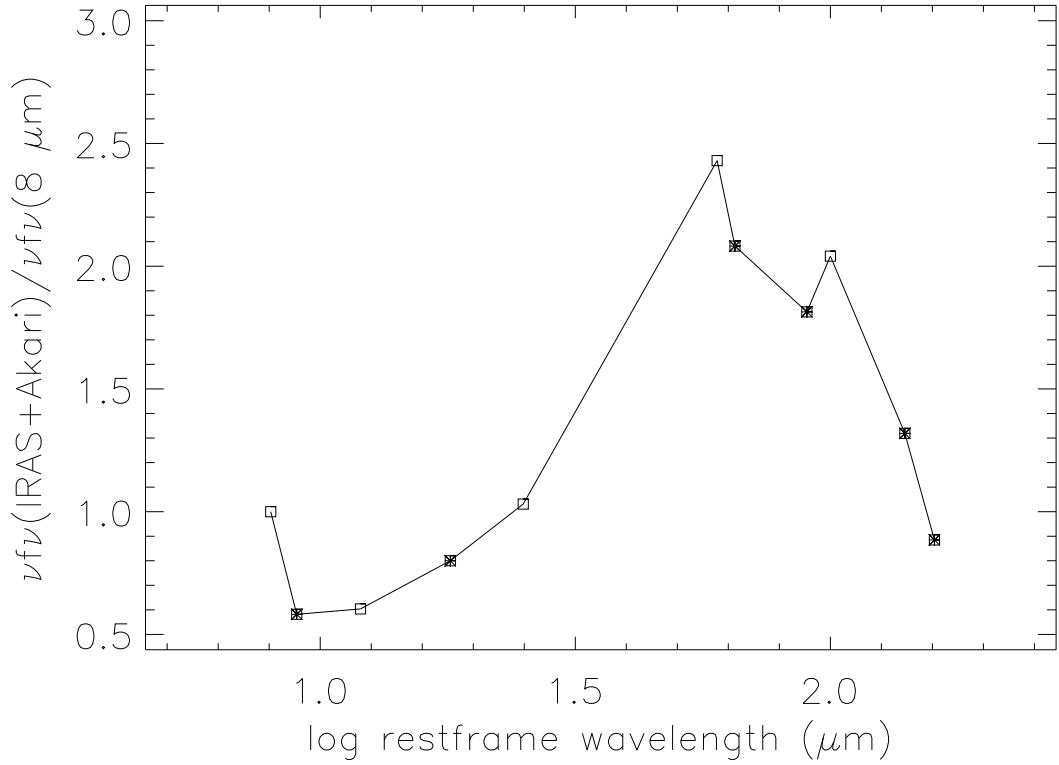}
\caption{ Median spectral energy distribution in the rest frame normalized to the peak $\nu$$f_{\nu}$($\sim$ 8 \ums) for all 190 sources having both IRAS (open squares) and complete AKARI (filled squares) flux densities.  AGN, composites, and starbursts are all combined in this plot.}

\end{figure}

\begin{figure}
\figurenum{9}
\includegraphics[scale=0.9]{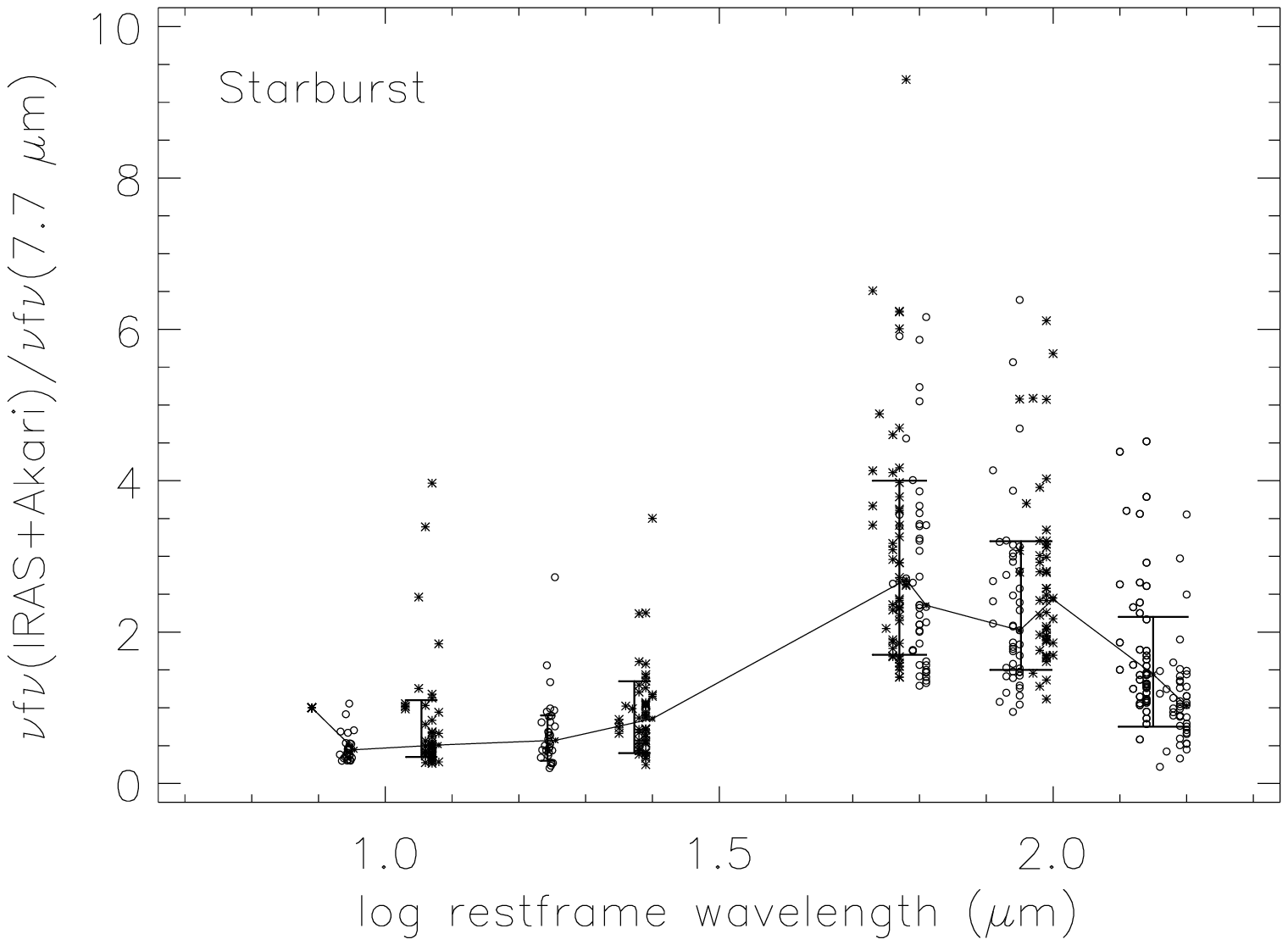}
\caption{ Spectral energy distributions for all starburst sources as determined from IRAS and AKARI flux densities, in the rest frame normalized to the peak $\nu$$f_{\nu}$(7.7 \ums).  Solid line shows medians at the observing wavelengths.   Asterisks are flux densities from IRAS, and circles are from AKARI.  Observed points are at rest frame wavelengths and are distributed in wavelength because of different source redshifts.  Error bars show one $\sigma$ dispersions at the various rest frame wavelengths; dispersions are determined using combined points for 60 \um and 65 \ums, for 90 \um and 100 \ums, and for 140 \um and 160 \ums.  }

\end{figure}


\begin{figure}
\figurenum{10}
\includegraphics[scale=0.9]{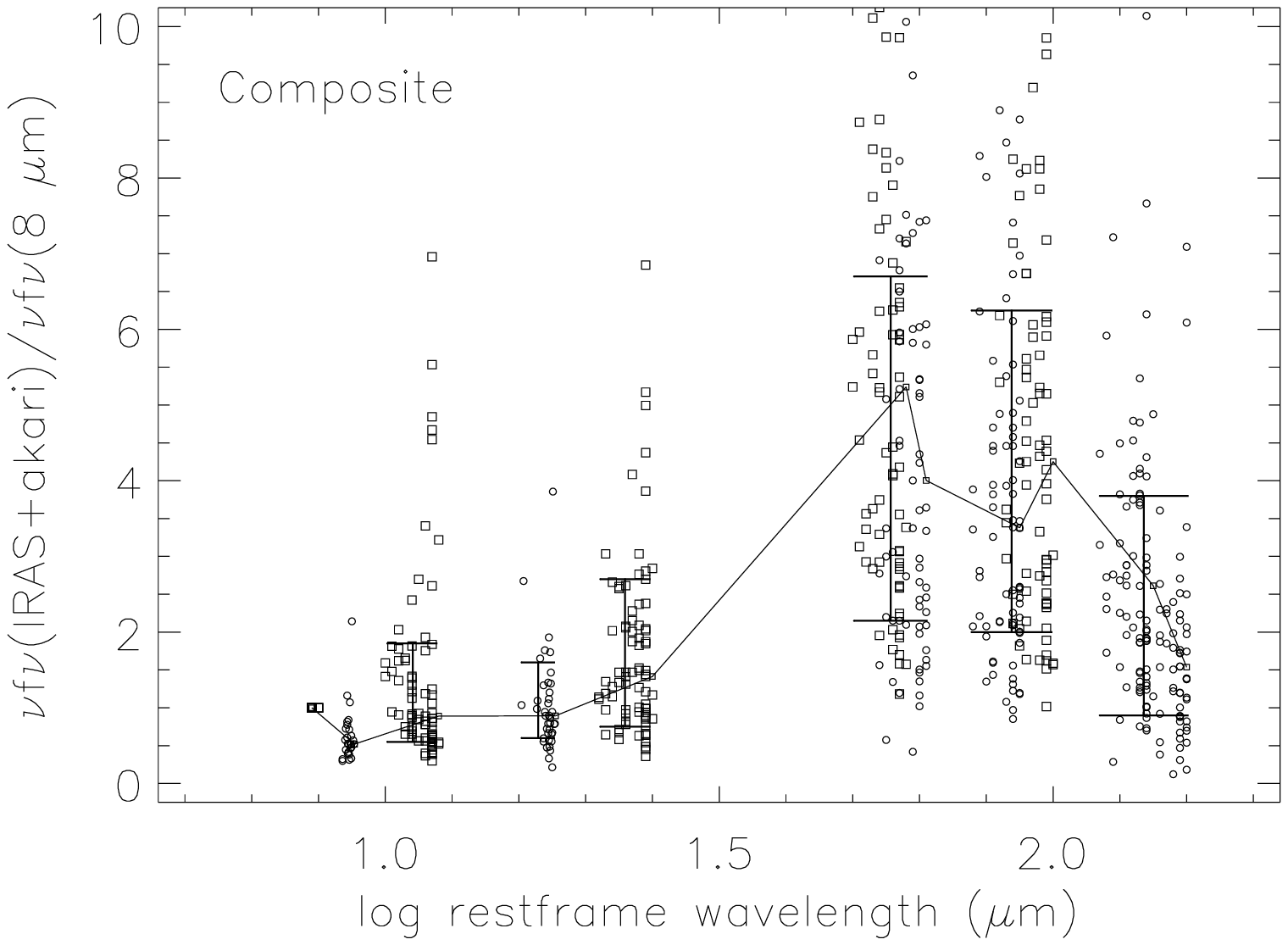}
\caption{ Spectral energy distributions for all composite sources as determined from IRAS and AKARI flux densities, in the rest frame normalized to the peak $\nu$$f_{\nu}$($\sim$8 \ums).  Solid line shows medians at the observing wavelengths. Squares are flux densities from IRAS, and circles are from AKARI.  Observed points are at rest frame wavelengths and are distributed in wavelength because of different source redshifts.  Error bars show one $\sigma$ dispersions at the various rest frame wavelengths; dispersions are determined using combined points for 60 \um and 65 \ums, for 90 \um and 100 \ums, and for 140 \um and 160 \ums.  }

\end{figure}


\begin{figure}
\figurenum{11}
\includegraphics[scale=0.9]{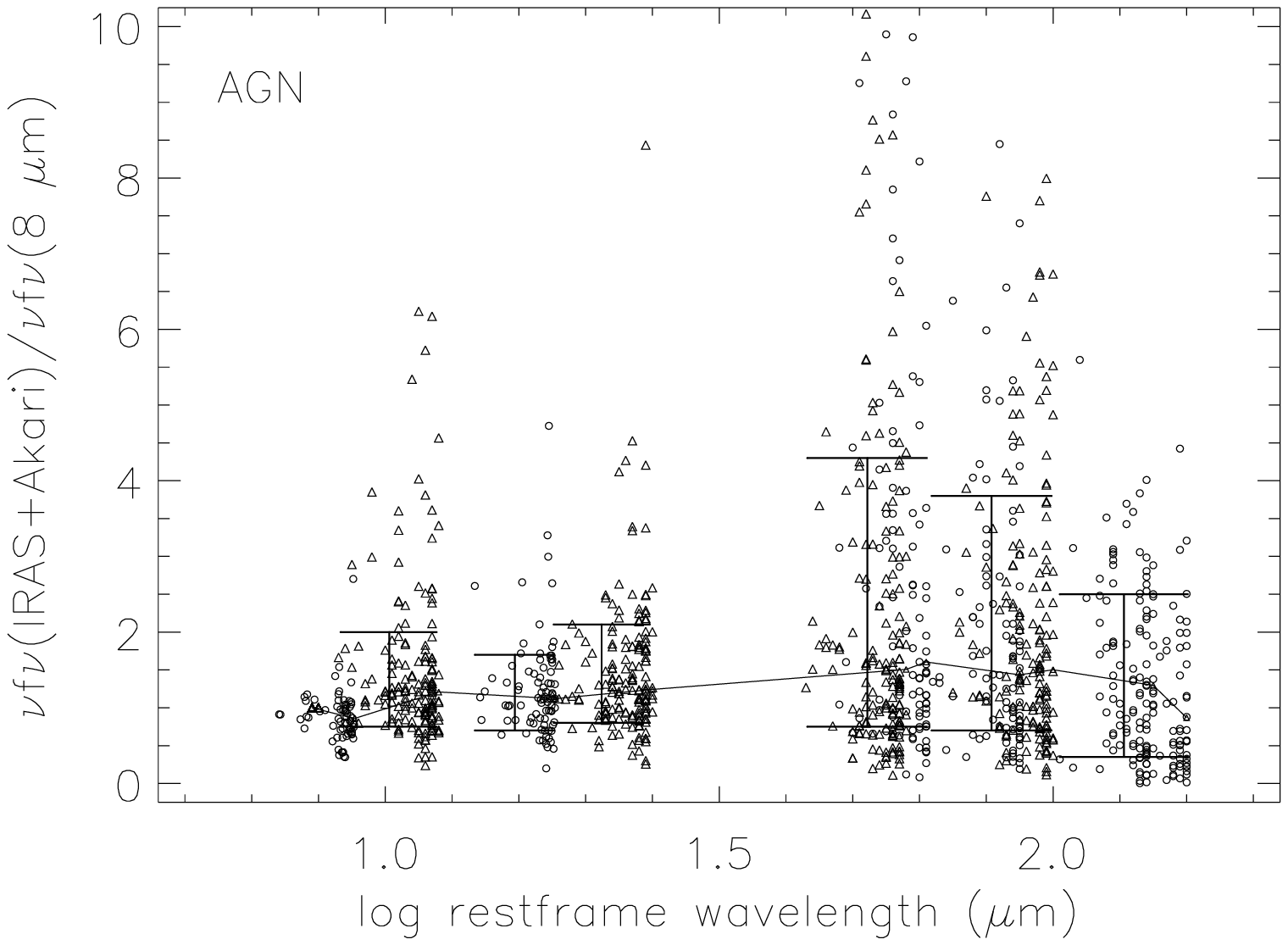}
\caption{ Spectral energy distributions for all AGN sources as determined from IRAS and AKARI flux densities, in the rest frame normalized to the peak $\nu$$f_{\nu}$($\sim$8 \ums).  Solid line shows medians at the observing wavelengths. Triangles are flux densities from IRAS, and circles are from AKARI.  Observed points are at rest frame wavelengths and are distributed in wavelength because of different source redshifts.  Error bars show one $\sigma$ dispersions at the various rest frame wavelengths; dispersions are determined using combined points for 60 \um and 65 \ums, for 90 \um and 100 \ums, and for 140 \um and 160 \ums. }

\end{figure}

\begin{figure}
\figurenum{12}
\includegraphics[scale=0.9]{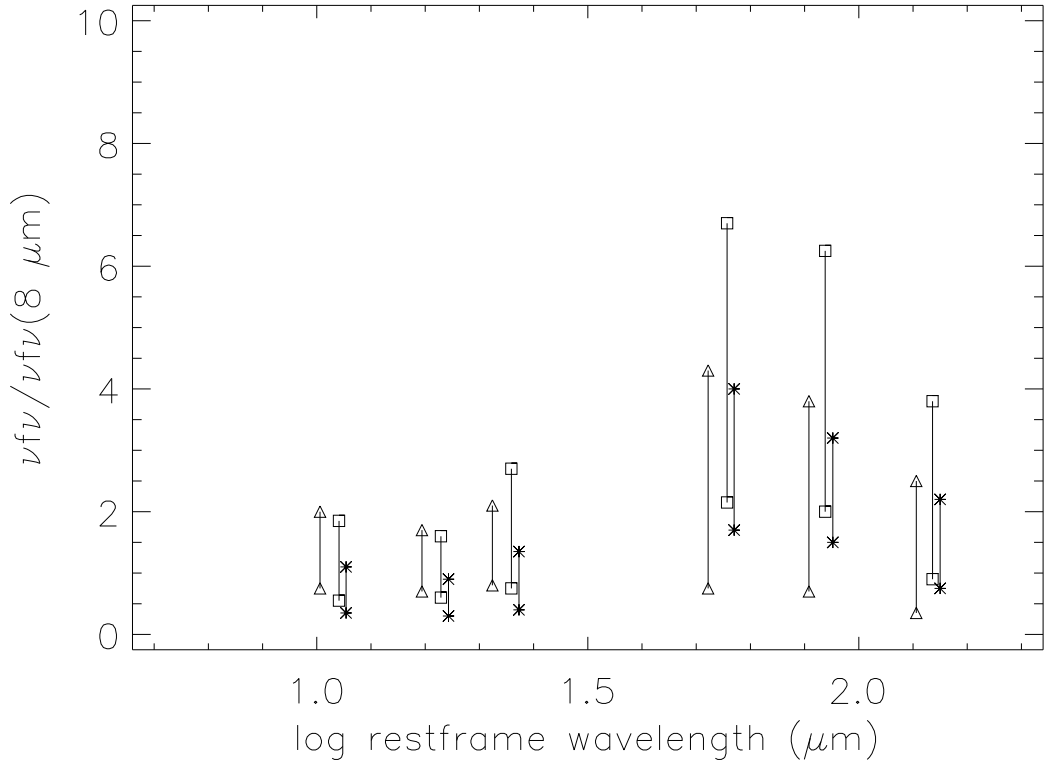}
\caption{ Dispersions for SEDs normalized to the peak $\nu$$f_{\nu}$($\sim$8 \ums) from Figures 9, 10, and 11 showing one $\sigma$ dispersions at the various rest frame wavelengths.  Asterisks are starbursts, squares are composites and triangles are AGN.}

\end{figure}

\begin{figure}
\figurenum{13}
\includegraphics[scale=0.9]{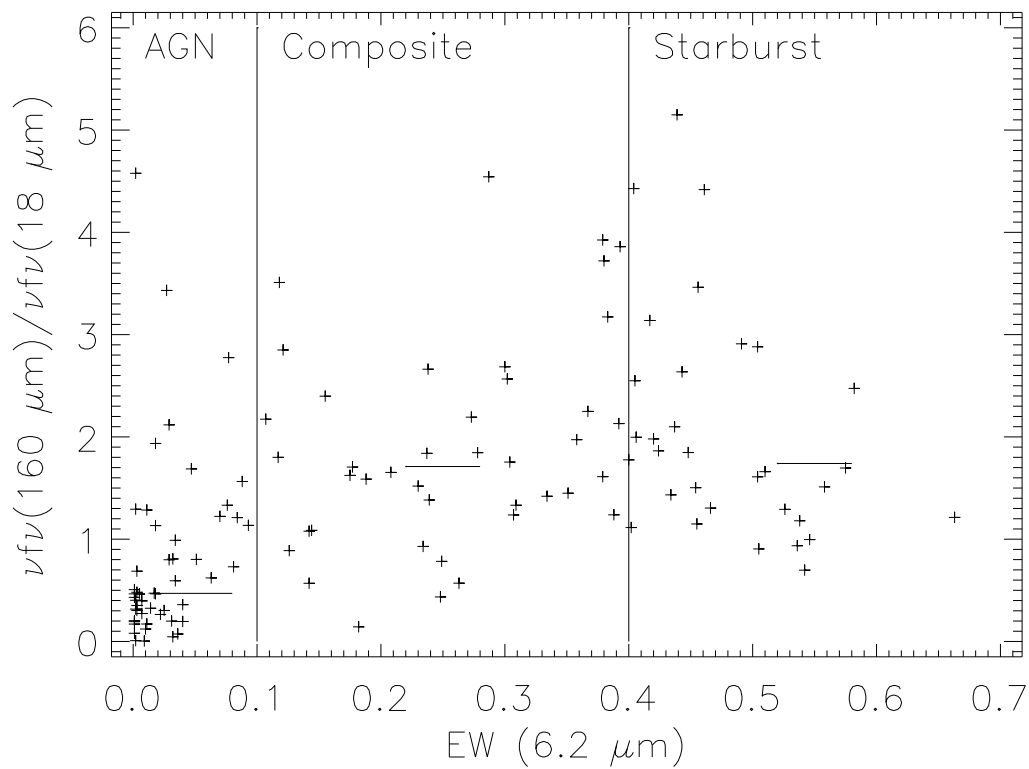}
\caption{ Comparison of cool dust component as measured by $\nu$$f_{\nu}$(160 \ums)/$\nu$$f_{\nu}$(18 \ums) from AKARI fluxes with strength of starburst as measured by PAH 6.2 \um feature.  Flux ratio shown is measure of relative luminosity in cooler dust radiating at 160 \um compared to warmer dust at 18 \ums.  Quantitative change in ratio shows how starbursts have systematically cooler dust than AGN; median ratio for starbursts is 1.74, median for composites is 1.71, and median for AGN is 0.47, but the lower limit of this ratio for pure starbursts (0.7) is greater than the median for AGN (0.47).}

\end{figure}

\begin{figure}
\figurenum{14}
\includegraphics[scale=0.9]{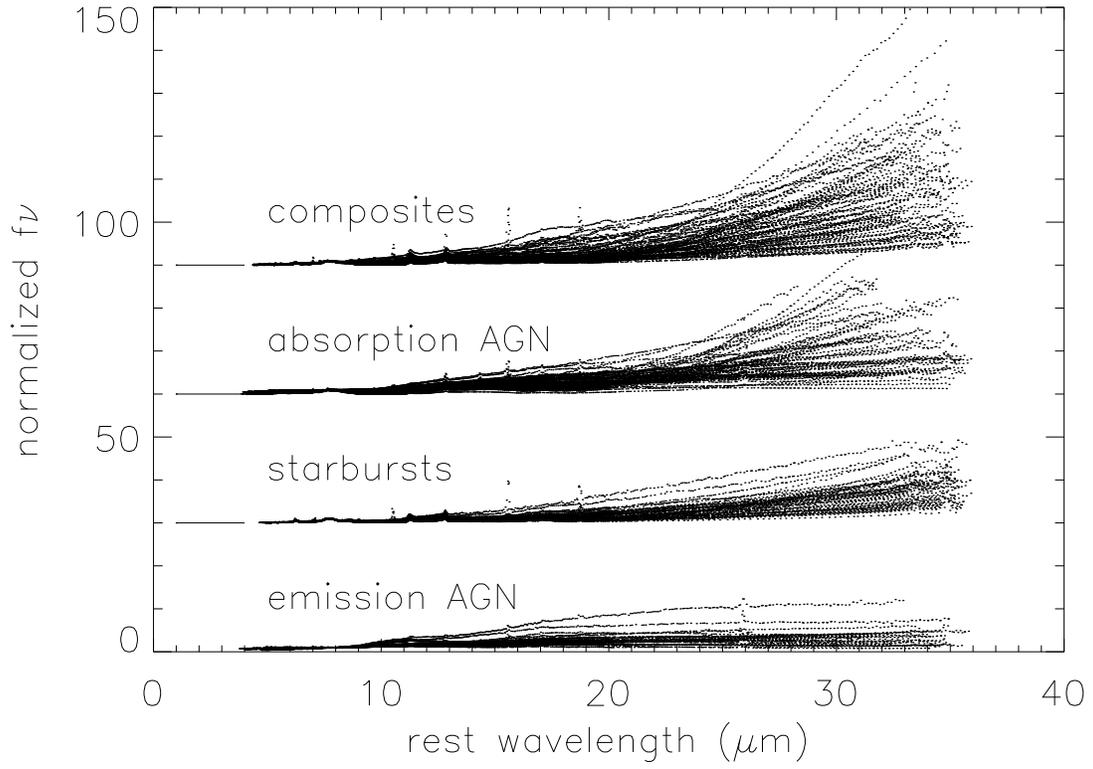}
\caption{Spectra for all sources included in Tables 1-6, normalized to the peak $f_{\nu}$($\sim$8 \ums), with individual sources overlaid to show dispersion and extremes in rest frame spectra among the different categories. Zero levels of spectra are shown by short horizontal lines.}

\end{figure}

\begin{figure}
\figurenum{15}
\includegraphics[scale=0.9]{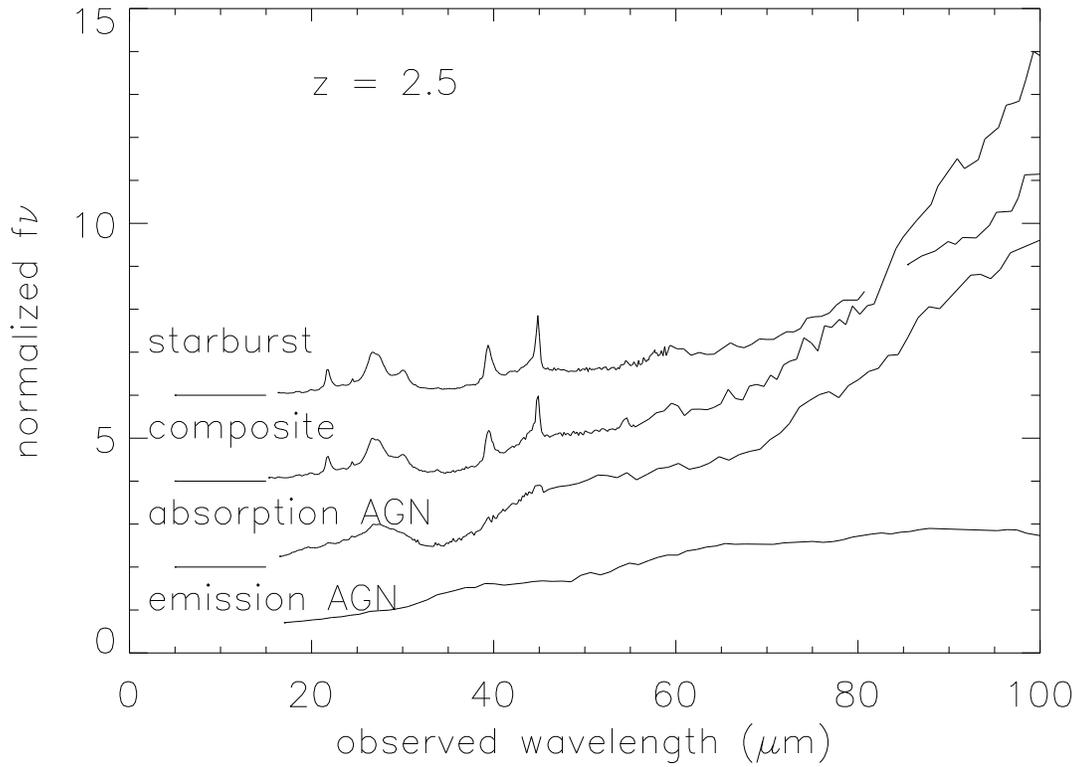}
\caption{Average spectra for sources of different classifications in Tables 1-6, normalized to the peak $f_{\nu}$($\sim$8 \ums), as would appear in the observed frame if redshifted to z = 2.5. Spectra are offset by two units of $f_{\nu}$ for illustration.  Zero levels of each spectrum are shown by short horizontal lines.  }

\end{figure}


\begin{thebibliography}


\bibitem[Armus et al.(2007)]{arm07}
Armus, L. et al. 2007, \apj, 656, 148
\bibitem[Bouwens et al.(2009)]{bou09}
Bouwens, R. J. et al. 2009, \apj, 705, 936
\bibitem[Brand et al.(2008a)]{brn08a}
Brand, K. et al. 2008a, \apj, 673, 119 
\bibitem[Brand et al.(2008b)]{brn08b}
Brand, K. et al. 2008b, \apj, 680, 119 
\bibitem[Brandl et al.(2006)]{bra06}
Brandl, B. et al. 2006, \apj, 653, 1129.
\bibitem[Calzetti(2008)]{cal08}
Calzetti, D. 2008, Astronomical Society of the Pacific Conference Series, 390, 121
\bibitem[Chapman et al.(2010)]{cha10}
Chapman, S. C. et al. 2010, \mnras, 409, L13
\bibitem[Dasyra et al.(2009)]{das09} 
Dasyra, K. M. et al. 2009, \apj, 701, 1123
\bibitem[Desai et al.(2009)]{des09}
Desai, V. et al. 2009, \apj, 700, 1190
\bibitem[Farrah et al.(2007)]{far07}
Farrah, D. et al. 2007, \apj, 667, 149 
\bibitem[Farrah et al.(2008)]{far08}
Farrah, D. et al. 2008, \apj, 677, 957 
\bibitem[Farrah et al.(2009a)]{far09a}
Farrah, D., Weedman, D., Lonsdale, C.J., Polletta, M., Rowan-Robinson, M., Houck, J., and Smith, H.E. 2009, \apj, 696, 2044 
\bibitem[Farrah et al.(2009b)]{far09b}
Farrah, D. et al. 2009b, \apj, 700, 395
\bibitem[Genzel et al.(1998)]{gen98}	
Genzel, R. et al. 1998, \apj, 498, 579
\bibitem[Gunn et al.(1998)]{gun98}
Gunn, J. E., et al. 1998, \aj, 116, 3040
\bibitem[Griffin et al.(2010)]{gri10}
Griffin, M. et al. 2010, \aap, 518, 3
\bibitem[Hao et al.(2005)]{hao05}
Hao, L. et al. 2005, \apjl, 625, L75
\bibitem[Hernan-Caballero et al.(2009)]{her09}
Hernan-Caballero, A. et al. 2009, \mnras, 395, 1695
\bibitem[Higdon et al.(2004)]{hig04}
Higdon, S.J.U. et al. 2004, \pasp, 116, 975
\bibitem[Houck et al.(2004)]{hou04} 
Houck, J. R. et al. 2004, \apjs, 154, 18
\bibitem[Houck et al.(2005)]{hou05}
Houck, J.R. et al. 2005, \apjl, 622, L105 
\bibitem[Houck et al.(2007)]{hou07} 
Houck, J. R., Weedman, D.W., LeFloc'h, E., and Hao, L. 2007, \apj, 671, 323
\bibitem[Houck and Weedman(2010)]{hou10}
Houck, J.R, and Weedman, D.W, 2010, in press, "Reionization to Exoplanets: $Spitzer's$ Growing Legacy", ASP Conference Series, ed: P. Ogle, arXiv1005.0344H
\bibitem[Hovhannisyan et al.(2010)]{hov10}
Hovhannisyan, A., Sargsyan, L. A., Mickaelian, A. M., and Weedman, D. W., Astrofizika, in press
\bibitem[Haung et al.(2009)]{hua09}
Huang, J. S. et al. 2009, \apj, 700, 183
\bibitem[Hunt et al.(2010)]{hun10}
Hunt, L. K, Thuan, T. X., Izotov, Y. I., and Sauvage, M. 2010, \apj, 712, 164. 
\bibitem[Imanishi et al.(2007)]{ima07}
Imanishi, M., Dudley, C. C., Maiolino, R., Maloney, P. R., Nakagawa, T., and Risaliti, G. 2007, \apj, 171, 72
\bibitem[Kaneda et al.(2007)]{kan07}
Kaneda, H., Kim, W., Onaka, T., Wada, T., Ita, Y., Sakon, I., and Takagi, T. 2007, \pasj, 59, S423
\bibitem[Kawada et al.(2007)]{kaw07}
Kawada, M., et al. 2007, \pasj, 59, S389
\bibitem[Kennicutt(1998)]{ken98}
Kennicutt, R.C. 1998, \araa, 36, 18
\bibitem[Lebouteiller et al.(2010)]{leb10}
Lebouteiller, V., Bernard-Salas, H., Sloan, G. C., and Barry, D. J. 2010, \pasp, 122, 231
\bibitem[Madau et al.(1998)]{mad98}
Madau, P., Pozzetti, L., and Dickinson, M. 1998, \apj, 498, 106
\bibitem[Markwick-Kemper et al.(2007)]{mk07}
Markwick-Kemper, F., Gallagher, S. C., Hines, D. C., and Bouwman, J. 2007, \apjl, 668, L107
\bibitem[Martinez-Sansigre et al.(2008)]{mar08}
Martinez-Sansigre, A., Lacy, M., Sajina, A., and Rawlings, S. 2008, \apj, 674, 676
\bibitem[Menendez-Delmestre et al.(2009)]{men09}
Menendez-Delmestre, K. et al. 2009, \apj, 699, 667
\bibitem[Murakami et al.(2007)]{mur07}
Murakami, H., et al. 2007, \pasj, 59, S369
\bibitem[Oliver et al.(2010)]{oli10}
Oliver, S. et al. 2010, \aap, 518, L21 
\bibitem[Onaka et al.(2007)]{ona07}
Onaka, T., et al. 2007, \pasj, 59, S401
\bibitem[Peeters et al.(2004)]{pee04}
Peeters, E., Spoon, H.W.W., and Tielens, A.G.G.M. 2004, \apj, 613, 986
\bibitem[Pilbratt et al.(2010)]{pil10}
Pilbratt, G. et al. 2010, \aap, 518, 1
\bibitem[Polletta et al.(2008)]{pol08}
Polletta, M., Weedman, D., Honig, S., Lonsdale, C. J., Smith, H. E., and Houck, J. 2008, \apj, 675, 960
\bibitem[Pope et al.(2008)]{pop08}
Pope, A. et al., 2008, \apj, 575, 1171
\bibitem[Rowan-Robinson et al.(2010)]{rr10}
Rowan-Robinson, M. et al. 2010, \mnras, 409, L2
\bibitem[Reddy and Steidel(2009)]{red09}
Reddy, N. A. and Steidel, C. C. 2009, \apj, 692, 778
\bibitem [Rieke et al.(2004)]{rie04}
Rieke, G.H. et al., 2004, \apjs, 154, 25
\bibitem[Sajina et al.(2007)]{saj07}
Sajina, A., Yan, L., Armus, L., Choi, P., Fadda, D., Helou, G., and Spoon, H. 2007, \apj, 664, 713  
\bibitem[Sanders and Mirabel (1996)]{san96}
Sanders, D. B., and Mirabel, I. F. 1996, \araa, 34, 749
\bibitem[Sargsyan et al.(2008)]{sar08}
Sargsyan, L., Mickaelian, A., Weedman, D., and Houck, J. 2008, \apj, 683, 114
\bibitem[Sargsyan and Weedman(2009)]{sar09}
Sargsyan, L. A. and Weedman, D. W. 2009, \apj, 701, 139
\bibitem[Sargsyan et al.(2010)]{sar10}
Sargsyan, L. A., Weedman, D. W.,and Houck, J. R. 2010, \apj, 715, 986
\bibitem[Schweitzer et al.(2008)]{sch07}
Schweitzer, M. et al. 2008, \apj, 679, 101.
\bibitem[Shi et al.(2006)]{shi06}
Shi, Y. et al. 2006, \apj, 653, 127 
\bibitem[Skrutskie et al.(2006)]{skr06}
Skrutskie, M. F. et al. 2006, \aj, 131, 1163
\bibitem[Smith et al.(2007)]{smi07}
Smith, J. D. T. et al. 2007, \apj, 656, 770
\bibitem[Soifer, Neugebauer, and Houck(1987)]{soi87}
Soifer, B. T., Neugebauer, G., and Houck, J. R. 1987, \araa, 25, 187
\bibitem[Spoon et al.(2007)]{spo07}
Spoon, H. W. W. et al. 2007, \apjl, 654, L49 
\bibitem[Weedman et al.(2006a)]{wee06a}
Weedman, D.W., Le Floc'h, E., Higdon, S.J.U., Higdon, J.L., and Houck, J.R. 2006a, \apj, 638, 613
\bibitem[Weedman et al.(2006b)]{wee06b}
Weedman, D.W. et al. 2006b, \apj, 653, 101.
\bibitem[Weedman and Houck(2008)]{wee08}
Weedman, D.W. and Houck, J.R. 2008, \apj, 686, 127 
\bibitem[Weedman and Houck(2009a)]{wee09a} 
Weedman, D.W. and Houck, J.R. 2009a, \apj, 693, 370
\bibitem[Weedman and Houck(2009b)]{wee09b}
Weedman, D.W. and Houck, J.R. 2009b, \apj, 698, 1682 
\bibitem[Wright(2006)]{wri06}
Wright, E. L. 2006, \pasp, 118, 1711
\bibitem[Wu et al.(2010)]{wu10}
Wu, Y. et al. 2010, \apj, 723, 895
\bibitem[Yan et al.(2007)]{yan07}
Yan, L. et al. 2007, \apj, 658, 778






\end{thebibliography}
\end{document}